%% file: main.tex
\documentclass[a4paper,11pt,english]{article}
\usepackage{jinstpub} 
\usepackage[utf8]{inputenc} 
\usepackage[T5,T1]{fontenc}
\usepackage[capitalize]{cleveref}
\usepackage{url}            
\usepackage{booktabs}       
\usepackage[group-separator={,},group-minimum-digits = 4]{siunitx}
\usepackage{nicefrac}       
\usepackage[protrusion=true,
            expansion=true]{microtype}      
\usepackage[caption=false]{subfig}
\usepackage{dcolumn}
\usepackage{microtype}
\usepackage{listings}
\usepackage[useregional]{datetime2}
\usepackage[export]{adjustbox}
\usepackage{lineno}
\usepackage{bm}

\newcommand{\bvec}[1]{\bm{\mathbf{#1}}}

\newcommand{\cth}{\cos \theta}

\newcommand{\reff}{r_A}
\newcommand{\rmod}{r_F}
\newcommand{\tmod}{t_F}

\newcommand{\zen}{\Theta_{\text{s}}}
\newcommand{\zeni}{\Theta_{\text{s},i}}
\newcommand{\hatzen}{\hat{\Theta}_{\text{s}}}
\newcommand{\hatzs}{\hat{Z}_{\text{s}}}
\newcommand{\zenr}{\Theta_{\text{r}}}
\newcommand{\xs}{X_{\text{s}}}
\newcommand{\ys}{Y_{\text{s}}}
\newcommand{\zs}{Z_{\text{s}}}
\newcommand{\bvecXs}{\bvec{X}_\text{s}}
\newcommand{\zsj}{Z_{\text{s},j}}
\newcommand{\cthzen}{\cos \zen}

\newcommand{\azi}{\Phi_{\text{s}}}
\newcommand{\edp}{E_{\text{dep}}}
\newcommand{\hatedp}{\hat{E}_{\text{dep}}}
\newcommand{\tres}{t_{\text{res}}}
\newcommand{\Pref}{\mathcal{P}}
\newcommand{\zref}{z_\mathcal{P}}
\newcommand{\zice}{z_\mathcal{I}}
\newcommand{\iot}[1]{\iota^{(#1)}}
\newcommand{\eqflat}[1]{
    A \Bigl( \reff \bigl(\bvec{r}_z, #1 \bigr), \theta, \phi, \zen, #1 \Bigr) F \Bigl( \rmod \bigl(\bvec{r}_z, #1 \bigr), \theta, \phi, \tres-\tmod \bigl(\bvec{r}_z, #1 \bigr), \zen, #1 \Bigr)}

\title{\boldmath Improved modeling of in-ice particle showers for IceCube event reconstruction}

\collaboration{
IceCube collaboration}

\input{authors}

\emailAdd{analysis@icecube.wisc.edu}

\abstract{The IceCube Neutrino Observatory relies on an array of photomultiplier tubes to detect Cherenkov light produced by charged particles in the South Pole ice. IceCube data analyses depend on an in-depth characterization of the glacial ice, and on novel approaches in event reconstruction that utilize fast approximations of photoelectron yields. Here, a more accurate model is derived for event reconstruction that better captures our current knowledge of ice optical properties. When evaluated on a Monte Carlo simulation set, the median angular resolution for in-ice particle showers improves by over a factor of three compared to a reconstruction based on a simplified model of the ice. The most substantial improvement is obtained when including effects of birefringence due to the polycrystalline structure of the ice. When evaluated on data classified as particle showers in the high-energy starting events sample, a significantly improved description of the events is observed.
}

\keywords{Cherenkov detectors, Neutrino detectors, Simulation methods and programs}

\arxivnumber{2403.02470} 

\begin{document}
\maketitle
\flushbottom

\section{Introduction}
\label{sec:intro}

The IceCube Neutrino Observatory detects neutrinos interacting with nucleons and electrons in the South Pole ice via Cherenkov radiation produced by charged secondaries. The ice is instrumented with \num{5160} digital optical modules (DOMs), each with a single downward-facing photomultiplier tube (PMT), arrayed across a cubic-kilometer volume below the surface. The DOMs are attached to 86 strings---cables installed in the ice that provide mechanical and electrical support. The DOMs are vertically spaced \SI{17}{\m} apart on standard IceCube strings and \SI{7}{\m} apart on DeepCore strings, a denser infill region of the detector. Standard IceCube strings are spaced approximately \SI{125}{\m} apart~\cite{Aartsen:2016nxy}.

IceCube covers a rich and diverse physics program at a wide energy range, and can detect neutrinos with energies spanning from about \SI{5}{\giga \eV} to above \SI{10}{\peta \eV}, as well as bursts of \si{\mega \eV} neutrinos from sufficiently nearby sources. Highlights include the discovery and subsequent confirmation of a diffuse flux of astrophysical neutrinos~\cite{IceCube:2013low,IceCube:2020wum}, the first identification of an astrophysical neutrino source, TXS 0506+056~\cite{IceCube:2018dnn}, arising from IceCube's real-time program~\cite{IceCube:2016cqr}, and the detection of the first astrophysical neutrino interaction at the Glashow resonance~\cite{IceCube:2021rpz}. More recently, the nearby Seyfert galaxy NGC 1068~\cite{IceCube:2022der} and the Milky Way itself~\cite{IceCube:2023ame} have been identified as steady sources of astrophysical neutrinos. These and future results rely on, and will continue to benefit from, refined calibration of the detector and improved event reconstruction.

In IceCube there are, broadly speaking, three general event categories: tracks, cascades, and double cascades. In this paper, these terms refer to the hypothesized model underlying event reconstructions, which is related to but conceptually distinct from the actual physical process of particle showers or muon propagation through matter~\cite{Workman:2022ynf}. High-energy muons produced in muon neutrino charged-current (CC) interactions can travel large distances through the detector and appear track-like. Electromagnetic (EM) particle showers induced by electron neutrino CC interactions or hadronic showers from neutral-current (NC) interactions of all flavors appear as cascades, a roughly spherical deposition of light in the detector. Tau neutrino CC interactions can appear as a double cascade for certain tau lepton decay channels and energies. For the purposes of event reconstruction, high-energy muons can be approximated as a series of cascades due to muon stochastic losses~\cite{Aartsen:2013vja}, and other events can be approximated using a single- or two-cascade model~\cite{hallen2013measurement}.

\begin{figure}[hbt]
\centering
\includegraphics[width=1.\linewidth]{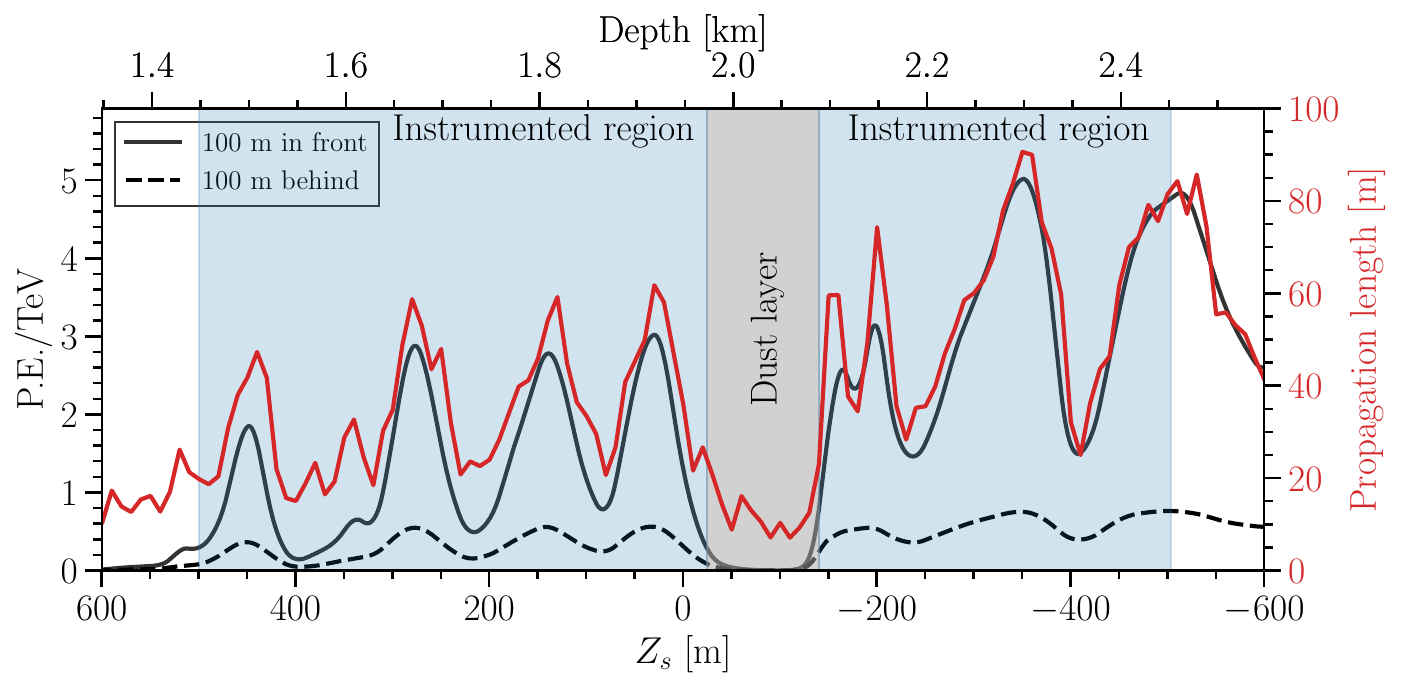}
\caption{Time-integrated photoelectron (P.E.) yields for a particle shower traveling horizontally ($\zen = 90^\circ$) at depth $\zs$, given in both detector coordinates (lower axis) and surface depth (upper axis), are shown in black. The solid (dashed) black line indicates the number of expected photoelectrons for receivers placed \SI{100}{\m} in front of (behind) the shower. The red line shows the depth dependence of a \SI{400}{\nano \m} photon's propagation length (right axis), which is proportional to the geometric mean of the absorption length and effective scattering length as defined in Ref.~\cite{Ackermann:2006pva}. Since the ice is a natural medium formed over epochs, large variations in optical properties that appear over large commensurate time periods are observed~\cite{IceCube:2013jrb}. A region of heightened dust concentration is highlighted in gray. The instrumented depths that IceCube spans are highlighted in blue, and include the dust layer. Clear differences for the two receivers illustrate how shower directionality can be reconstructed, while the strong correlation between yield and propagation length highlights the importance of accurate ice modeling.}
\label{fig:yield}
\end{figure}

The glacial ice serves as a natural Cherenkov medium for detection of the byproducts of neutrino interactions. From the surface to the bedrock, ice isochrons have formed over geological time scales, each with different scattering and absorption lengths that affect the propagation of Cherenkov photons. Calibration LEDs on board each DOM have been used to provide a detailed measurement of these optical properties as a function of depth~\cite{IceCube:2013llx}. The propagation length for a typical Cherenkov photon at \SI{400}{\nano \m} is shown in red in \cref{fig:yield}, and illustrates the depth-dependent optical properties of the natural ice sheet. The propagation length is defined in Ref.~\cite{Ackermann:2006pva}, and is proportional to the geometric mean of the absorption length and effective scattering length. For this illustration, perfectly flat ice isochrons are assumed and no additional ice anisotropies are present besides the depth-dependent scattering and absorption. In reality, we now know that the ice is more complex and exhibits additional anisotropies. These include an axial dependence now attributed to birefrigence of the polycrystals in the ice, which leads to a higher photon yield along the axis of glacial ice flow~\cite{tc-18-75-2024}. Elevation changes of ice isochrons, or ice layer undulations, have also been mapped out in detail throughout the detector volume~\cite{IceCube:2023qua}. Inclusion of these effects into a model usable for event reconstruction is our primary focus.

Accurate event reconstruction requires accurate modeling of photoelectron arrival time distributions for a given physics hypothesis. In this paper we focus on energies most relevant for high-energy astrophysical neutrinos, above roughly \SI{10}{\tera \eV}. As an example, \cref{fig:yield} provides a visualization of the expected Cherenkov photoelectron yield arising from an EM shower. The shower is placed at various depths in the detector, $\zs$, with $\zs=0$ corresponding to a depth of \SI{1948}{\m} below the surface, arriving from a direction $\zen = 90^\circ$ relative to vertical. Photoelectron yields are shown for receivers, which are idealized representations of IceCube's downward facing PMTs, placed at distances \SI{100}{\m} in front of (black, solid) and behind (black, dashed) the shower. Note the impact of the propagation length (red) on the photoelecton yields.  Differences between yields at the two receivers illustrate how shower directionality can be reconstructed, while the strong correlation to photon propagation length highlights the importance of accurate ice modeling.

The rest of this paper is organized as follows. In \cref{sec:cha}, we provide an overview of some of the challenging aspects of shower reconstruction in IceCube. \Cref{sec:bspl} describes improvements when fitting tabulated photoelectron distributions from Monte Carlo (MC) simulations with tensor-product B-splines, which reduce observed artifacts in the reconstructed zenith distribution while improving the median angular resolution for a benchmark simulation set by about $1^\circ$ at \SI{1}{\peta \eV} from an original resolution of \SI{12}{\degree}. \Cref{sec:bfr} details an effective treatment to include ice birefringence without increasing the dimensionality of the model used in reconstruction, further lowering the median angular resolution by $5.3^\circ$ at \SI{1}{\peta \eV}. \Cref{sec:tilt} describes a correction to account for ice layer undulations that yields another $1.7^\circ$ improvement at \SI{1}{\peta \eV}. \Cref{sec:considerata} touches on an approximation of shower longitudinal extensions with a two-cascade model, which yields an additional $0.5^\circ$ improvement at \SI{1}{\peta \eV}, resulting in a final median angular resolution of \SI{3.5}{\degree}, and includes a brief discussion of systematic uncertainties. In \cref{sec:summary}, we provide a summary of the results and conclude the discussion.

\section{Challenges in shower reconstruction}
\label{sec:cha}
\begin{figure}[hbt]
\centering
\includegraphics[height=0.5\linewidth]{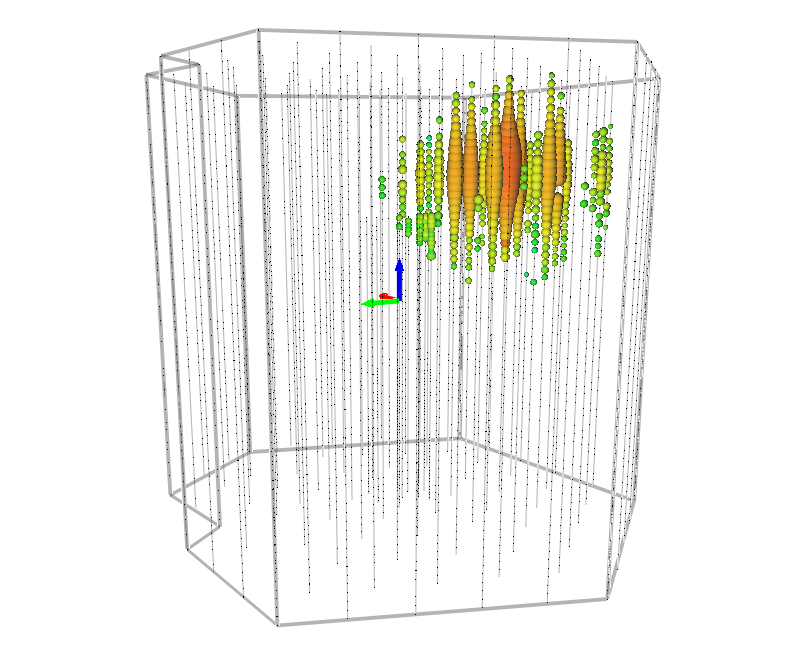}
\caption{Event view of an approximately \SI{2}{\peta \eV} particle shower detected by IceCube on \DTMdisplaydate{2012}{12}{04}{-1}. Each string is shown as a thin line extending from top to bottom; small dots correspond to unhit DOMs. Modules that detected a photoelectron are indicated by the colored spheres. The size of each sphere corresponds to the total charge detected, and its color indicates the timing of hits, with earlier to later going from red to green. The IceCube-centered coordinate axes are indicated by the three colored arrows, with the blue arrow indicating the positive $z$ direction. The length of each arrow is set to \SI{100}{\m} and gives a sense of scale. Note that noise cleaning has been applied for visualization purposes~\cite{Aartsen:2016nxy}.}
\label{fig:shovel}
\end{figure}

\Cref{fig:shovel} shows an event view of a high-energy shower in IceCube. For events contained within a fiducial volume such as this, the energy reconstruction is relatively well constrained by calorimetry~\cite{Aartsen:2013vja}. Further, at the energies relevant in IceCube, particles produced by neutrino interactions in the ice are, to good approximation, colinear with the neutrino direction, and thus their reconstruction points to the neutrino arrival direction. However, due to the large length scales between IceCube sensors, the directional reconstruction of showers can be a challenge and depends critically on the ice model. Anisotropies in the ice can induce differences in detected Cherenkov photoelectron yields that mimic those caused by a shift in direction.

The most realistic model of photon propagation in the ice requires a full MC simulation. The parallel nature of photon propagation makes such a task well suited for graphics processing units (GPUs)~\cite{Chirkin:2013tma,Chirkin:2019rcj}. Such an approach has allowed for a refined description of the optical properties of the ice by simulating LED calibration devices and fitting to the observed in situ calibration data~\cite{IceCube:2013llx,tc-18-75-2024}. Physics simulations of particle showers also rely on MC-based photon propagation.

Processing speeds on GPUs may be sufficient for simulation, but event reconstruction often requires testing orders of magnitude more points in the physics parameter space for each event. As of this writing, utilizing GPUs for event reconstructions where each Cherenkov photon is fully resimulated for every tested hypothesis is only feasible for $O(100)$ events~\cite{Chirkin:2013avz,IceCube:2023sov}. At the expense of some accuracy, fast approximations of Cherenkov photon yields have been developed to address this. Initial approximations relied on look-up tables~\cite{Lundberg:2007mf}, which were improved to allow for gradient-based minimization using a tensor product of B-splines~\cite{Aartsen:2013vja, Whitehorn:2013nh}. More recently, neural network (NN)~\cite{IceCube:2021umt,Abbasi:2021ryj,IceCube:2023avo} models have been employed as well. All such models are approximations, being derived from full MC simulations of particle showers or minimum ionizing muons.

There are then two main challenges that arise in event reconstruction. The first is to accurately model the physics and instrumentation to the best extent possible within the full MC simulation chain. To a large extent this includes the modeling of the optical properties of the ice, where significant progress has been made over the past few years that led to an improved description of the observed ice anisotropy based on birefrigence~\cite{tc-18-75-2024} and a more realistic mapping of layer undulations~\cite{IceCube:2023qua}. Including these effects has resulted in much better agreement with in situ calibration data.

The second challenge is to quickly and accurately model the photoelectron yields derived from full MC simulation. Anisotropies in the ice due to birefringence and layer undulations bring further complexity by breaking symmetry and thus introducing additional dimensionality. A full parametrization of the photoelectron yield for a cascade would span nine dimensions: $(\xs,\ys,\zs)$ to define its position, $(\zen, \azi)$ to define its arrival direction, $(r, \theta, \phi)$ to define the DOM position relative to the cascade, and $t$, the photon arrival time at the DOM. This increased complexity makes a tensor product of B-splines intractable to evaluate over all relevant regions of the parameter space; memory requirements over the full nine dimensions would exceed \SI{500}{\tera B}. In addition, the simulation needed to construct approximators becomes prohibitively expensive to produce even on the Open Science Grid~\cite{Pordes:2007zzb}, likely exceeding \si{10^{10}} CPU hours. While NNs are not bound by these dimensionality constraints, they can be subject to unknown inaccuracies in interpolation and extrapolation while also requiring a large training dataset. Depending on the approach, for example if a fixed set of geometry, ice properties and detector settings are assumed in the simulation, the model might need to be retrained for each successive update~\cite{Abbasi:2021ryj,IceCube:2021umt,IceCube:2023avo}. Furthermore, training an NN may provide less insight into where limitations in simulation and reconstruction lie. If MC simulation is taken as an absolute ground truth, then we are subject to any existing mismodeling there. A more iterative approach, with cross checking between simulation and expected results from event reconstruction, can help resolve issues that exist either in the MC or in the construction of any approximate model.

\subsection{Impact of ice anisotropies on shower reconstruction}
\label{sec:impact}

\begin{figure}[hbt]
\centering
\includegraphics[width=0.49\linewidth]{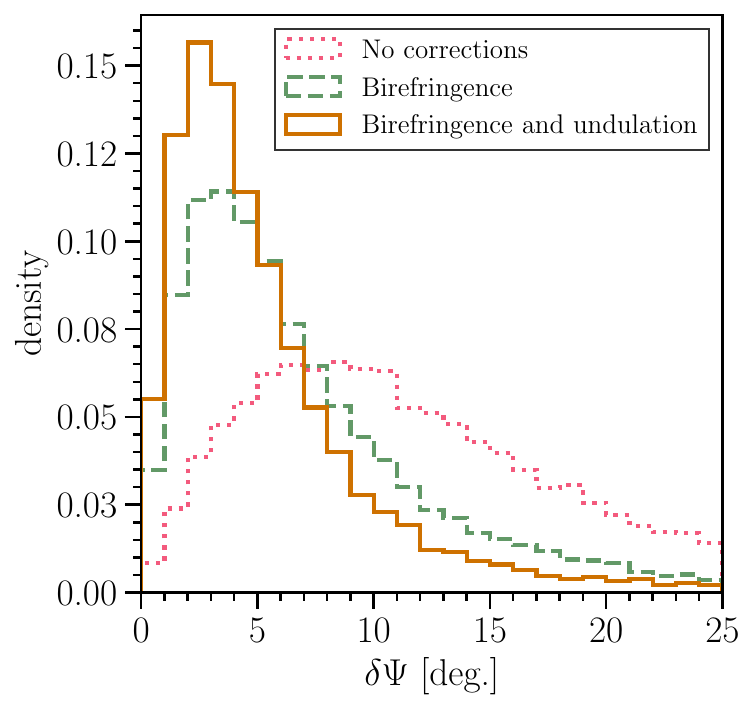}
\includegraphics[width=0.5\linewidth]{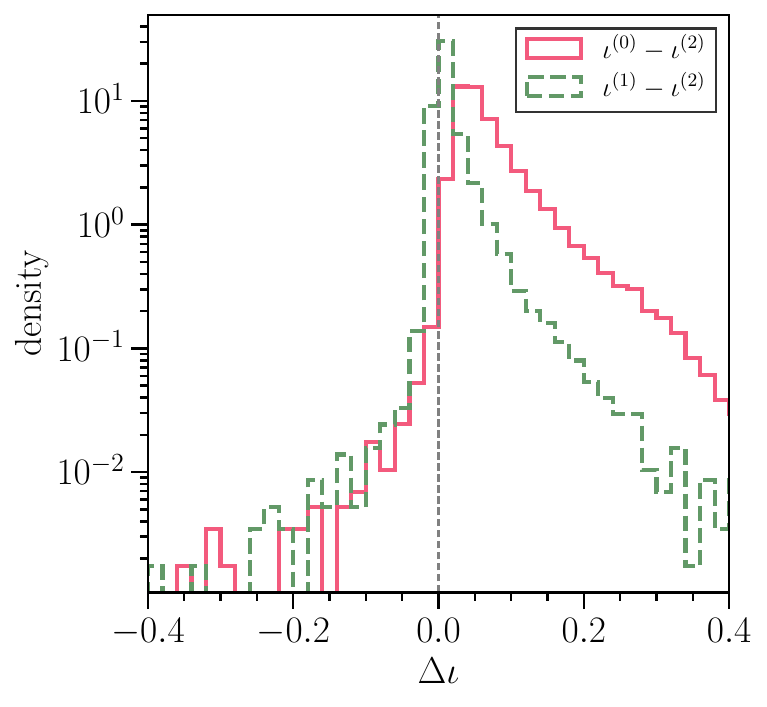}
\caption{These distributions highlight the improvements in $\delta \Psi$ and $\iota$ when corrections that approximate the discussed ice anisotropies are included in reconstruction. The left panel compares $\delta \Psi$ distributions with no corrections (red, dotted), with birefringence (green, dashed) and with both layer undulation and birefrigence (orange, solid). Each correction successively reduces the space angle between the true and reconstructed direction. The right panel shows two distributions across $\Delta \iota$ relative to a model that includes all corrections. Lower $\iota$ values correspond to improved agreement with simulated events in the benchmark MC. The reduced negative log-likelihood for a model that includes corrections due to both layer undulations and birefringence is given by $\iot{2}$, including only the birefringence correction $\iot{1}$, and including no corrections $\iot{0}$. The improved description with $\iot{2}$ is evidenced by the skew towards positive values, which is more apparent for $\iot{0} - \iot{2}$ (red, solid) but still visible for $\iot{1} - \iot{2}$ (green, dashed). The vertical dashed line at $\Delta \iota = 0$ serves as a visual guide to highlight the positive skew.}
\label{fig:iota}
\end{figure}

To confront the asymmetries introduced by a refined understanding of the ice, a set of corrections to approximate birefringence and layer undulations can be employed. As will be shown, these corrections do not increase the dimensionality of the model; six dimensions is enough, with $(\zen, \zs)$ to fix the zenith angle and depth of the shower and $(r, \theta, \phi, t)$ to define the arrival position and time at a receiving DOM~\cite{Lundberg:2007mf}. The impact of these corrections can be evaluated on a benchmark, high-energy starting events (HESE) MC set, consisting of simulated showers from electron-neutrino and electron-antineutrino interactions contained within a fiducial region of the detector~\cite{IceCube:2020wum}. Electron (anti)neutrinos are sampled from an $E^{-1.5}$ spectrum at the Earth's surface, propagated through the Earth to reach IceCube, whereupon they may interact and produce signatures that pass the HESE selection criteria. After cuts, the events range in energies from \SI{10}{\tera \eV} to \SI{10}{\peta \eV}, a regime where the astrophysical neutrino flux is expected to transition above the atmospheric neutrino background. The simulation relies on GPUs for photon propagation~\cite{Chirkin:2019rcj} and incorporates our most up-to-date knowledge of the ice, namely birefringence~\cite{tc-18-75-2024} and recent updates to ice layer undulations~\cite{IceCube:2023qua}. The left panel of \cref{fig:iota} shows the distribution of the angle between the true direction and reconstructed direction, $\delta \Psi$. Including the birefringence correction (green, dashed) improves upon the event reconstruction that employs no corrections (red, dotted). Including both birefrigence and layer undulation corrections (orange) exhibits even better performance. The implementation of these corrections is the subject of \cref{sec:bfr} and \cref{sec:tilt}.

Another metric for comparison is the negatived log-likelihood per degree of freedom (reduced negative log-likelihood), $\iota$, obtained by comparing expected photoelectron yields and time profiles from the best-fit cascade to the simulated data. Here we use an effective Poisson-based likelihood~\cite{Arguelles:2019izp}, modified to use a fixed relative uncertainty across all bins to capture residual error in the model. The right panel of \cref{fig:iota} illustrates the $\iota$ improvement when all corrections are applied. Shown are two $\Delta \iota$ distributions, $\iot{1} -\iot{2}$ and $\iot{0} - \iot{2}$, where $\iot{0}$, $\iot{1}$ and $\iot{2}$ refer to reduced negative log-likelihoods for event reconstructions that assume no corrections, birefrigence correction only, and all ice-associated corrections, respectively. As lower values correspond to an improved event description compared to the simulated data, the positive skew observed for both distributions indicates the improvement of $\iot{2}$ over $\iot{1}$ and $\iot{0}$.

\section{Improvements in model construction}
\label{sec:bspl}

\begin{figure}[hbt]
\centering
\includegraphics[width=0.6\linewidth]{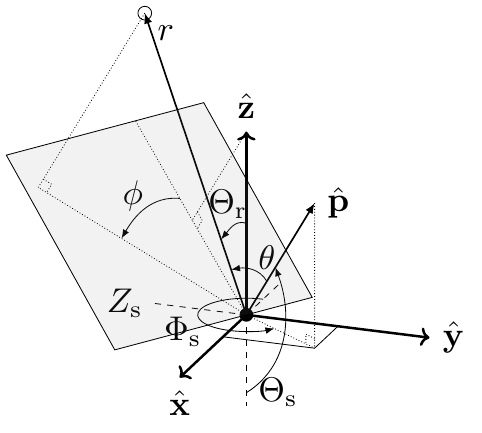}
\caption{The convention used to describe source and receiver coordinates~\cite{Lundberg:2007mf}. Source and receiver positions are indicated as filled and empty circles, respectively. The unit vector $\hat{\bvec{p}}$ denotes the shower momentum direction, while $(\zen, \azi)$ describes the arrival direction, $-\hat{\bvec{p}}$, in the standard spherical coordinate system with polar axis aligned with $\hat{\bvec{z}}$. The receiver coordinates, $(r, \theta, \phi)$, are given in a spherical coordinate system centered on the source position and defined such that the polar axis aligns with $\hat{\bvec{p}}$ and $\phi=0$ corresponds to where the projection of $\bvec{r}$ onto the plane perpendicular to $\hat{\bvec{p}}$ (shaded) is maximally vertical. In a media with symmetry about the $\hat{\bvec{z}}$ axis, $\azi$ is degenerate. If, additionally, photon emission is axially symmetric about $\hat{\bvec{p}}$, $\phi$ is degenerate with $-\phi$.}
\label{fig:coordinates}
\end{figure}

As mentioned in \cref{sec:cha}, the photoelectron-yield model that is the focus of this paper derives from GPU-based MC photon propagation~\cite{Chirkin:2019rcj}. The convention used to describe source and receiver coordinates, introduced in Ref.~\cite{Lundberg:2007mf}, is shown in \cref{fig:coordinates}. For directional sources of Cherenkov photons, a simplified MC where birefringence (see \cref{sec:bfr}) and ice layer undulations (see \cref{sec:tilt}) are not included is used to tabulate yields, so that azimuthal symmetry is preserved about the $\hat{\bvec{z}}$ axis and translational symmetry is preserved over $(\xs, \ys)$. Further, in ice with depth-dependent optical properties (see \cref{fig:yield}) the axial symmetry of Cherenkov photon emission about the shower axis, $\hat{\bvec{p}}$, which holds to good approximation for the colinear particles produced in high-energy particle showers, is reduced to a bilateral symmetry where $\phi$ is degenerate with $-\phi$. Thus, the source can be completely described by $(\zen, \zs)$, and the receiver $\phi$ ranges from \SIrange{0}{180}{\degree}.

A \SI{1}{\giga \eV} EM shower, the photon source, is repeatedly simulated along a grid of points, $(\zeni, \zsj)$, indexed by $i$ and $j$. For each source, the photoelectron yield for receiver $k$ at a location $\bvec{r}_k = (r_k, \theta_k, \phi_k)$ relative to the source is tabulated as a function of
\begin{equation}
\tres(r) \equiv t - nr/c,
\label{eq:tres}
\end{equation}
where $n$ is the group index of refraction in ice and $t$ is the absolute photon arrival time~\cite{Lundberg:2007mf}. The receiver position is defined using a spherical coordinate system aligned with the shower principal axis, as shown in \cref{fig:coordinates}. Once tabulated, the total number of photoelectrons, or amplitude, as well as the cumulative density function (CDF) in $\tres$ is fitted across $(\bvec{r}_k, [\tres])$ for each simulated cascade in $(\zeni, \zsj)$ using a tensor product of B-splines. Then each fitted surface over the receiver coordinates is stacked across $(\zeni, \zsj)$~\cite{Whitehorn:2013nh}, resulting in a function of the form
\begin{equation}\label{eq:bulk}
    A(\bvec{r}, \zen, \zs) F(\bvec{r}, \tres, \zen, \zs),
\end{equation}
where $A$ is the amplitude and $F$ is the CDF. As the number of Cherenkov photons scales linearly with shower energy~\cite{Aartsen:2013vja}, \cref{eq:bulk} gives, within a constant factor, the number of photoelectrons detected by a DOM for an EM shower. 

\subsection{Interpolation}
\label{sec:inter}

\begin{figure}[hbt]
\centering
\includegraphics[width=0.49\linewidth]{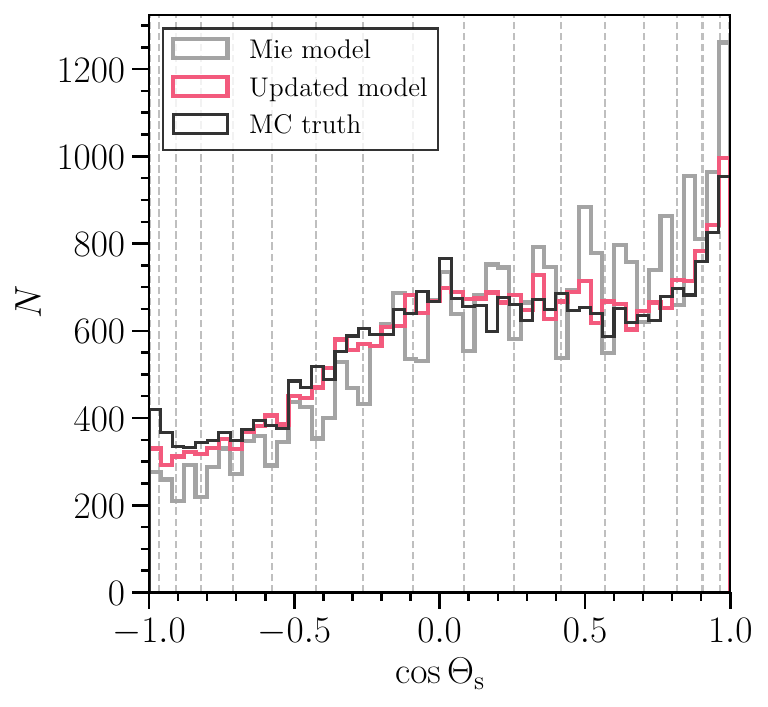}
\includegraphics[width=0.475\linewidth]{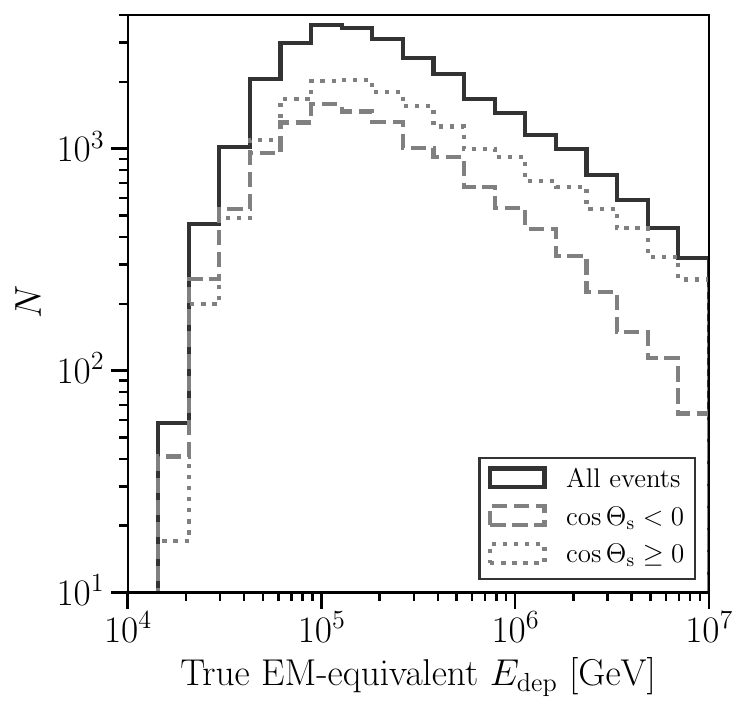}
\caption{The left panel shows the number of events per bin, $N$, distributed over $\cos \zen$ for the benchmark MC sample consisting of electron neutrino and antineutrino interactions contained within a fiducial volume~\cite{IceCube:2020wum}. The effects of birefrigence~\cite{tc-18-75-2024} and ice layer undulations~\cite{IceCube:2023qua} are included in the simulation, but are not included in either of the event reconstructions shown. A ringing effect is visible when the Mie model is used (gray), which disappears when the updates discussed in \cref{sec:bspl} are included (red). Vertical dashed lines indicate the placement of knot positions for the Mie B-splines along the $\cthzen$ dimension. The MC truth (black) is shown for comparison. To give a sense of the statistics and energies of this MC set, the right panel shows the overall $\edp$ distribution in black. Gray dashed and dotted lines show the breakdown for upgoing and downgoing events, respectively. See text for more details.}
\label{fig:recozenith}
\end{figure}

The first step towards improved modeling for shower reconstruction required resolving a long-standing issue that was observed in the reconstructed $\hatzen$ distribution. \Cref{fig:recozenith} illustrates the problem for our benchmark MC. The right panel shows MC statistics after applying HESE cuts~\cite{IceCube:2020wum} as a function of the true EM-equivalent deposited energy, $\edp$. Sometimes referred to as visible energy~\cite{Aartsen:2013vja}, $\edp$ is calculated for each simulated event by summing over the energies of all EM and hadronic shower components, with a correction that scales down the true energy of hadronic showers to match the energy of an EM shower that, on average, produces an equivalent number of Cherenkov photons. Gray dashed and dotted lines show the breakdown for upgoing and downgoing events, respectively. The left panel compares the true $\cthzen$ distribution (black) against the reconstructed distribution using a now-outdated model (gray)~\cite{Aartsen:2013vja}, which was derived from the best knowledge of the ice in 2013~\cite{IceCube:2013llx}. Following the terminology in Ref.~\cite{IceCube:2013llx}, we refer to this as the ``Mie model''. At the time, ice layers in the Mie model were assumed to be flat for purposes of event reconstruction, and no anisotropies beyond the depth dependence shown in \cref{fig:yield} were included. In the left panel, dashed vertical lines indicate B-spline knot positions in $\cthzen$ used in the Mie model, and a clear ringing pattern arising from an artificial pull away from knot locations emerges. The solution came in the form of improved interpolation with additional support points, the topic of this \lcnamecref{sec:inter} and illustrated by the $\cos \hatzen$ distribution in red.

Multidimensional B-spline surfaces constructed with a tensor product can have interpolation artifacts along diagonals across coordinate axes. This is due to defining basis functions on Cartesian grids~\cite{Khalil}. In the absence of dense support points, smooth features along diagonals are not preserved and interpolation exhibits an oscillatory behavior. In particular, this effect led to the ringing effect shown by the gray distribution in the left panel of \cref{fig:recozenith}. Including the updates described in this \lcnamecref{sec:bspl} results in the reconstructed $\cos \hatzen$ distribution shown in red, without ringing and in better agreement with the true $\cthzen$ distribution (black). The birefrigence and layer undulation corrections, which will be discussed in \cref{sec:bfr} and \cref{sec:tilt} respectively, were not included in the updated model (red) shown here. Including these effects would further improve agreement to the true distribution in the most upgoing region ($\cthzen \approx -1$).

\begin{figure}[hbt]
\centering
\includegraphics[width=0.49\linewidth]{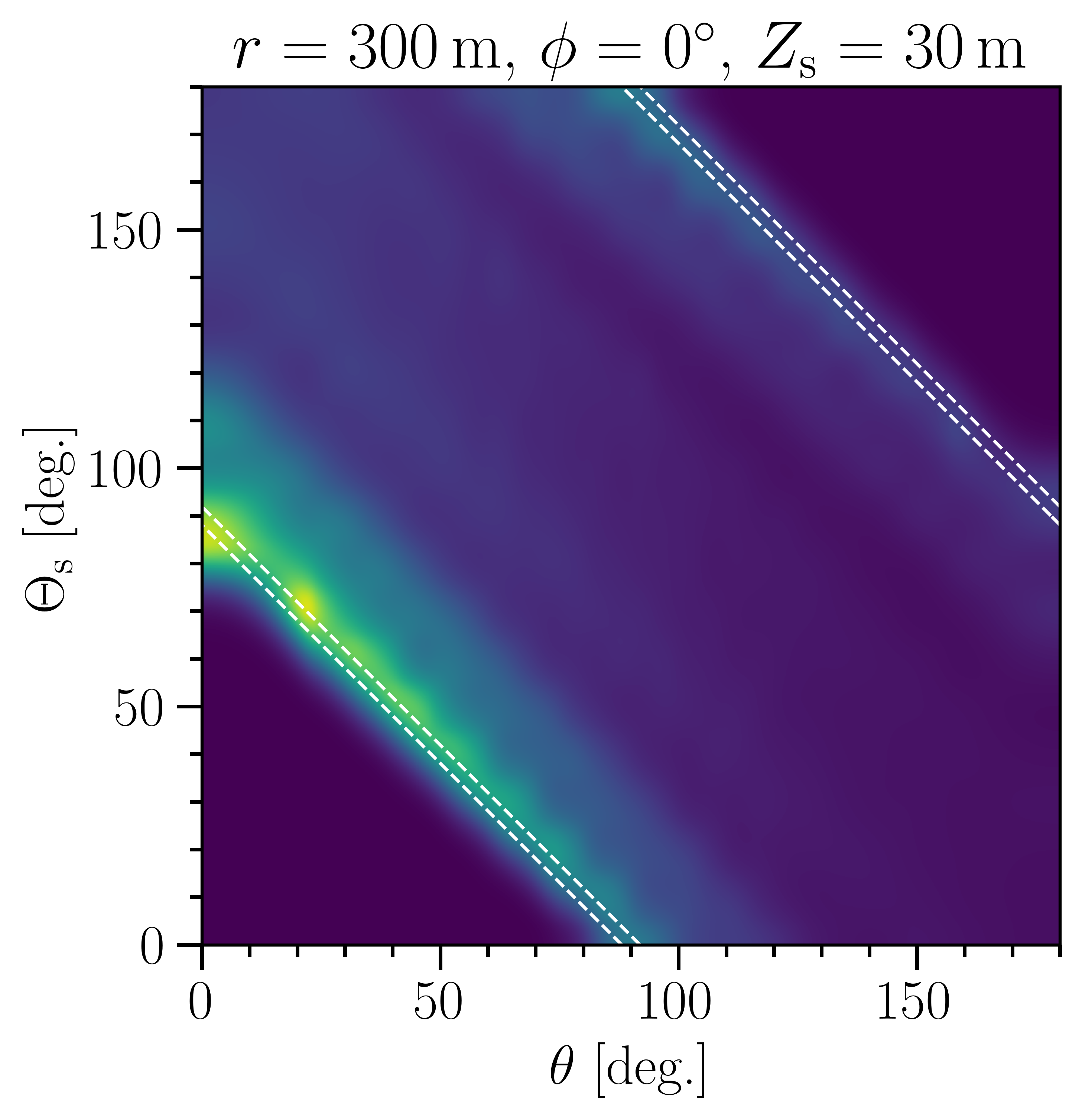}
\includegraphics[width=0.49\linewidth]{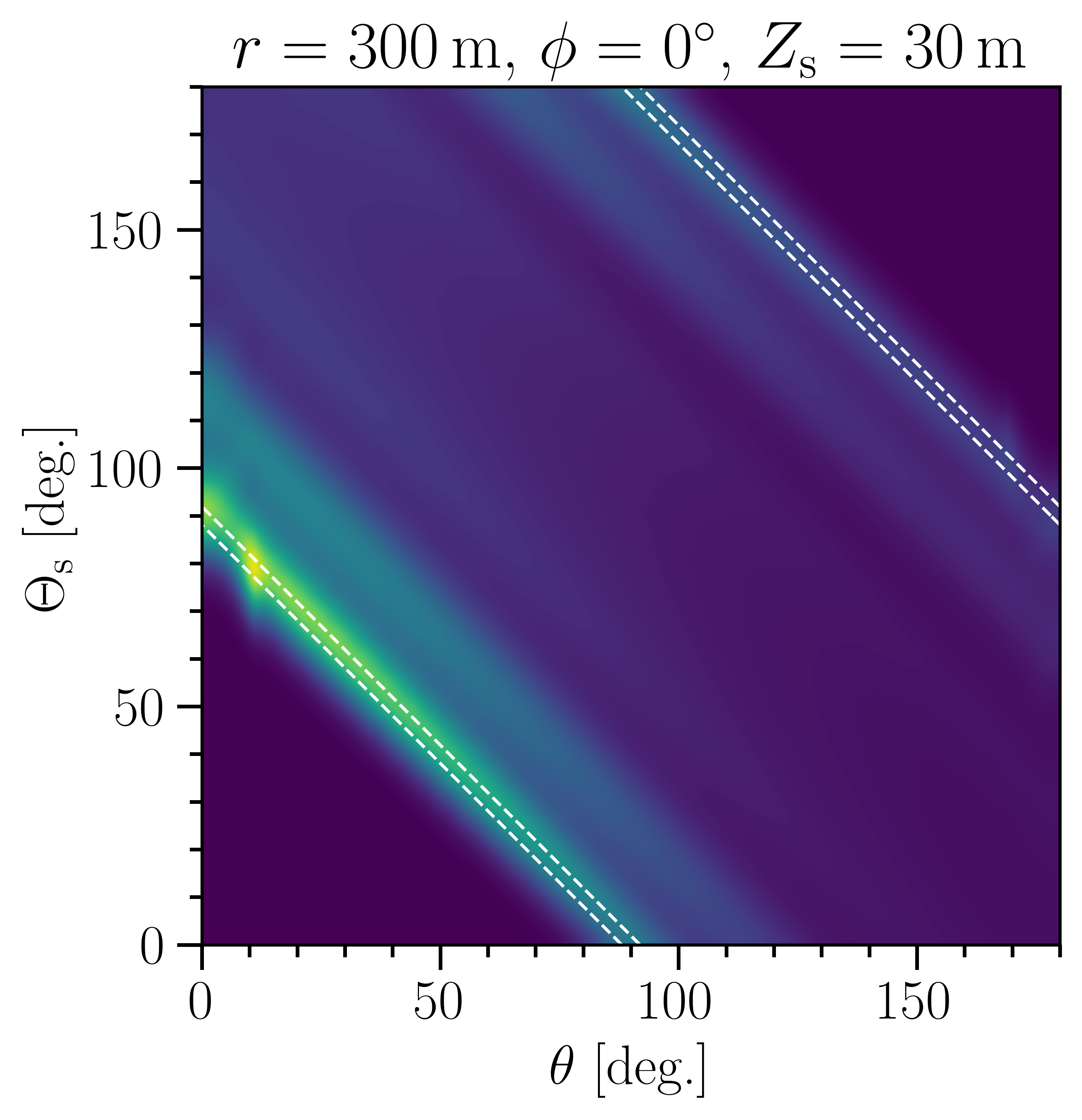}
\includegraphics[width=0.49\linewidth]{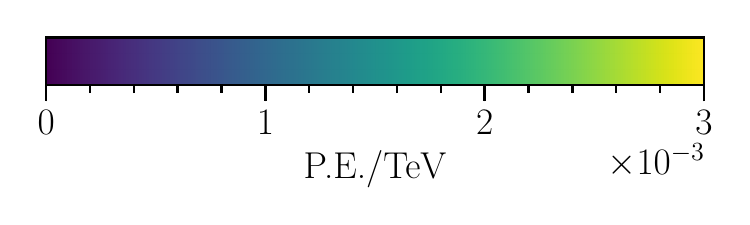}
\caption{Visualization of the time-integrated photoelectron (P.E.) yield as a function of $\theta$ and $\zen$, sliced across the other dimensions at the values indicated in the title. The left (right) panel shows the Mie (updated) model expectations. Dashed white lines bound regions in the phase space where receiver locations fall within $\pm \SI{10}{\m}$ of $\zs$ in depth, corresponding to a region of clear ice where the source is placed. The ringing in the left panel is due to the tensor product B-spline construction, which has difficulties interpolating along diagonals when support points are sparse. This is mitigated by increasing the number of support points in $\zen$, the result of which is shown in the right panel.}
\label{fig:thetazen}
\end{figure}
\Cref{fig:thetazen}, left panel, illustrates the interpolation artifacts observed in the amplitude, $A(\bvec{r}, \zen, \zs)$, for the Mie model at the coordinates given in the title. In this case, $\zs=\SI{30}{\m}$ corresponds to a source placed in a region of clear ice (see \cref{fig:yield}). As shown in \cref{fig:coordinates}, the receiver coordinate system is spherical and aligned with respect to $\zen$. For example, at $\zen=90^\circ$ a receiver at $\theta=0^\circ$ would be placed directly in front of the shower and hence in the same ice layer. As $\zen$ sweeps through its parameter space, a compensation in $\theta$ of equal magnitude ensures that the receiver is placed at the same depth as the source. The result is that for clear ice, the amplitude will be greater along that diagonal than away from it, where a photon would have traversed through regions of higher absorption and scattering. However, due to the low number of support points in $\zen$ (every $10^\circ$) over which the model is stacked, interpolation artifacts appear.

A simple, but effective, solution was to double the number of support points in $\zen$, tabulating at $5 ^\circ$ intervals instead of $10^\circ$. The number of knot locations in $\zen$ roughly doubles as well, the cause of which is described in Ref.~\cite{Whitehorn:2013nh}. With this change, the amplitude model yields smoother behavior along diagonals, as shown in the right panel of \cref{fig:thetazen}. Updated values of the depth-dependent ice optical properties are included in the simulation used to fit the model~\cite{IceCube:2023qua}. However, similar to the construction of the Mie model, the effects of birefringence and ice layer undulations are turned off for reasons of simplification discussed in \cref{sec:cha} and at the beginning of this \lcnamecref{sec:bspl}. In both panels, the dashed white lines bound regions in the phase space where receiver locations fall within $\pm \SI{10}{\m}$ of $\zs$ in depth. Note that while the visualization here is for the amplitude model, the denser support points are applied to the CDF model in $\tres$ as well.

\begin{figure}[hbt]
\centering
\includegraphics[width=0.49\linewidth]{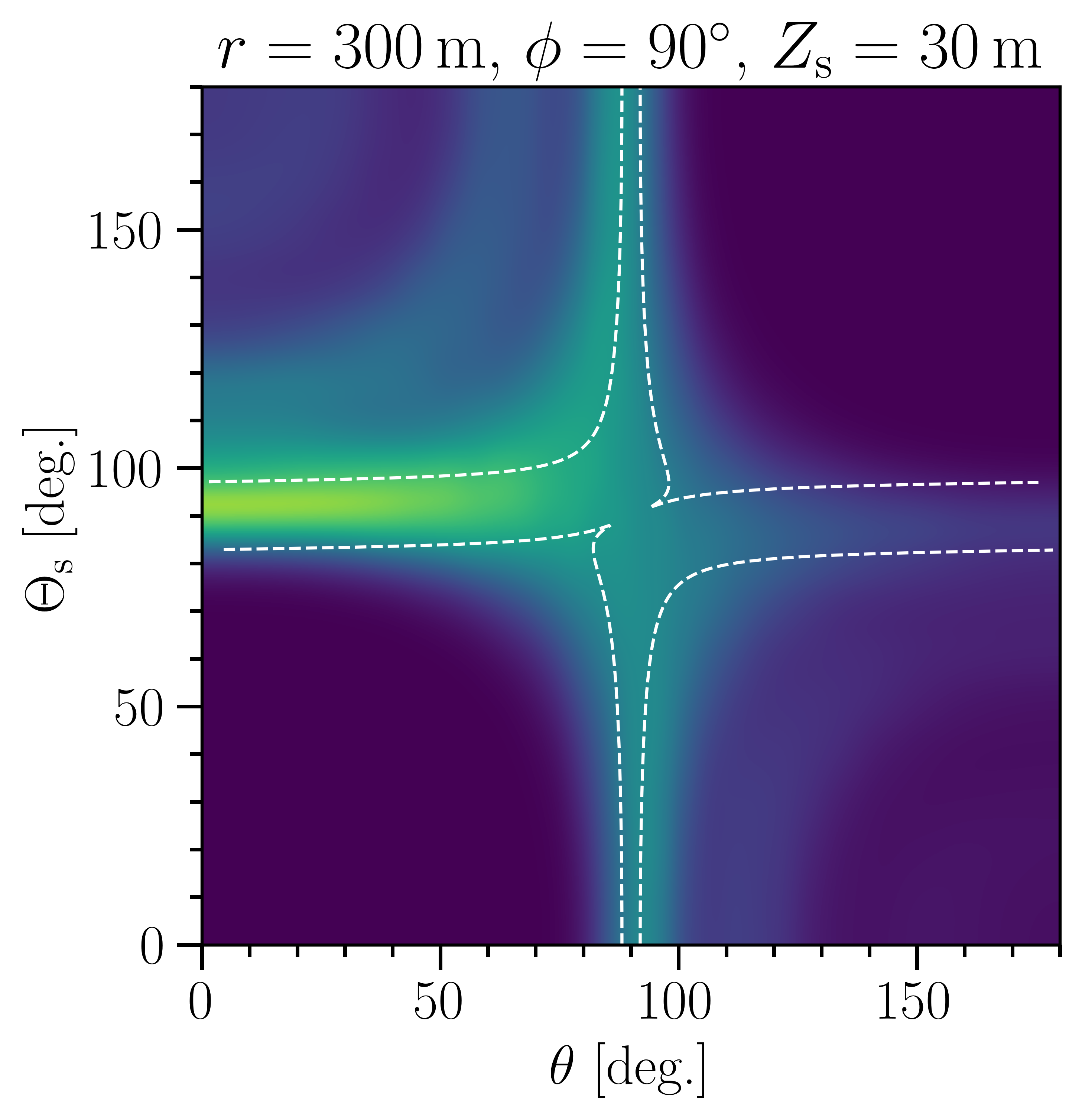}
\includegraphics[width=0.49\linewidth]{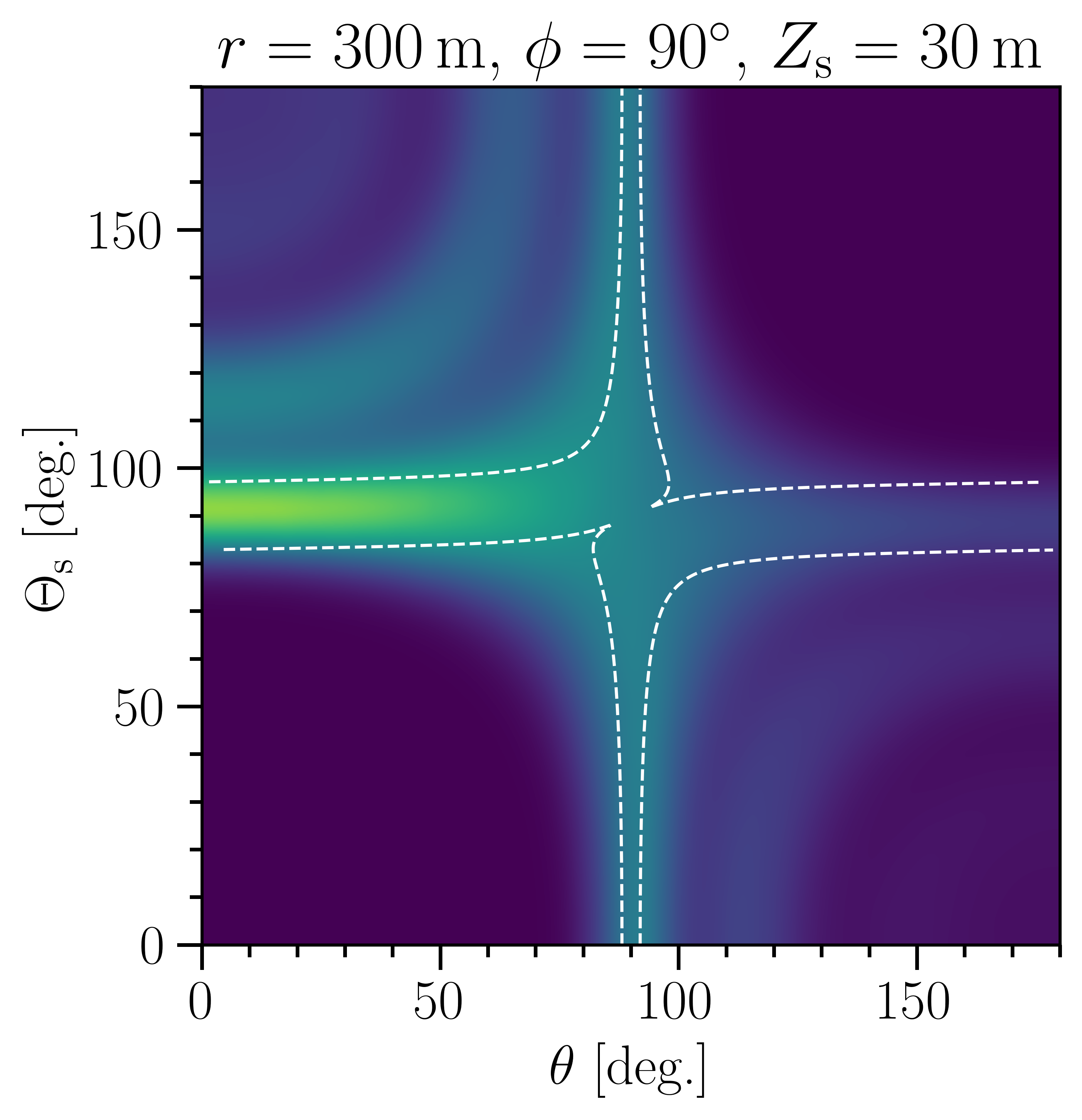}

\includegraphics[width=0.49\linewidth]{figures/theta_vs_zen_colorbar.pdf}
\caption{Same as \cref{fig:thetazen} except at a different slice, $\phi=90^\circ$. The dashed white lines bound regions in the phase space where receiver locations fall within $\pm \SI{10}{\m}$ of $\zs$ in depth, corresponding to a layer of clear ice where the source is placed. Here, large amplitude regions no longer lie along diagonals in $(\theta, \zen)$ space but instead align with the coordinate axes. As a result, interpolation artifacts do not appear in either the Mie model (left panel) or the updated model with denser support points (right panel).}
\label{fig:thetazenphi90}
\end{figure}
\Cref{fig:thetazen} only presents a single two-dimensional slice; other parameters are fixed at the coordinates indicated in the title. The full effect across all dimensions is more complex, and an example of what happens when $\phi = 90^\circ$ is shown in \cref{fig:thetazenphi90}. This also explains why, although a simple coordinate transformation could resolve the issues in the slice shown in \cref{fig:thetazen}, a generalized coordinate transformation over the additional $\phi$ dimension does not exist in simple form and hence we resorted to increasing the number of support points in $\zen$.

\subsection{Extrapolation}
\label{sec:extra}

Extrapolation failures can also occur when a parameter's domain extends beyond its last support point. These failures led to sharp discontinuities in the photoelectron yields across boundaries, and they occurred in the spherical coordinates system defined by $\theta$ and $\phi$, near the poles where $\cos \theta = \pm 1$ and near the boundary of $\phi \in \{0^\circ, 180^\circ\}$. The former required the construction of additional data points at $\cth = \pm 1$ based on linear interpolation, and the latter was resolved by reflection across the boundary in $\phi$. By including these additional support points, extrapolation artifacts were largely reduced.
\begin{figure}[hbt]
\centering
\includegraphics[width=0.49\linewidth]{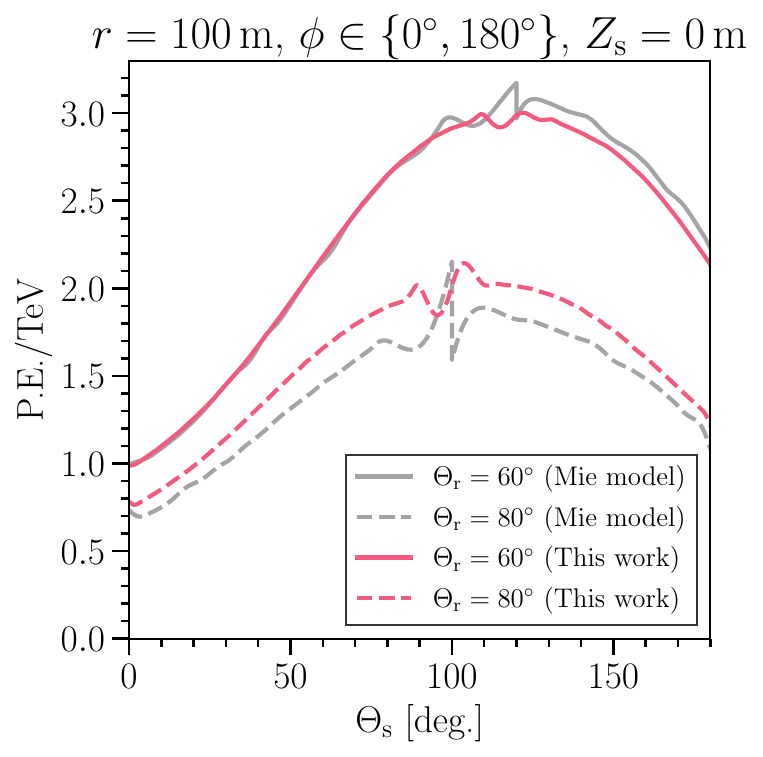}
\caption{The time-integrated photoelectron (P.E.) yield as a function of $\zen$ for two receivers placed at a fixed distance $r = \SI{100}{\m}$ from the source. The receivers are placed at two different directions, $\zenr$, of $\zenr=60^\circ$ (solid) and $\zenr=80^\circ$ (dashed), where $\zenr$ is the angle between source-receiver vector and $\hat{\bvec{z}}$ as shown in \cref{fig:coordinates}. A discontinuity is visible in the Mie model (gray), which is smoothed out with the updates discussed in this work (red). The interpolation artifacts are also visible as small fluctuations in the dashed lines, and are reduced by the updates discussed in \cref{sec:inter}.}
\label{fig:discont}
\end{figure}

The discontinuity at $\cth = \pm 1$ is again related to the tensor product construction of the model, which does not map naturally onto a spherical coordinate system. Simulation data is linearly binned in $\cth$, with the abscissa of each data point taken at the bin centers. Thus, at exactly $\cth=\pm 1$ no data point can be tabulated. To reach $\cth = \pm 1$, the Mie model extrapolated beyond the closest bin center. Furthermore, at $\cth = \pm 1$ the model would extrapolate to different values as a function of $\phi$, leading to the discontinuity visible in the gray lines in \cref{fig:discont}, which shows the time integrated amplitude as a function of $\zen$ for two receivers placed at $\zenr = \SI{60}{\degree}$ and \SI{80}{\degree}. Here, we have chosen $\bvec{r}$ to lie in the same plane as $\hat{\bvec{p}}$ and $\hat{\bvec{z}}$ for simplification, but these examples are representative of behavior across other dimensional slices. As shown in \cref{fig:coordinates}, when all three vectors lie in the same plane, $\theta = |\SI{180}{\degree}- \zen-\zenr|$ and
\begin{equation}
    \phi = 
    \begin{cases}
      0^\circ, & \zenr < 180^\circ - \zen\\
      180^\circ, & \zenr > 180^\circ - \zen.
    \end{cases}
\end{equation}
In a spherical coordinate system, the azimuthal angle $\phi$ becomes degenerate at the poles, and there the model prediction should not change as a function of $\phi$. Unfortunately, this cannot be guaranteed in the tensor product B-spline construction. The extrapolation in the Mie model, combined with the flip in $\phi$ when $\theta = 0^\circ$, cause the discontinuities in the gray lines in \cref{fig:discont}. 

The discontinuities can be largely reduced by the addition of a set of support points at $\cth = \pm 1$ such that the to-be-fitted data is the same across all $\phi$ at the poles. These data values were computed by first performing a linear interpolation of the tabulated data to $\phi=90^\circ$, then a linear extrapolation to $\cth = \pm 1$ along the $\phi = 90^\circ$ curve. By including these values in the B-spline fit, the polar discontinuity was reduced, as shown in the red lines in \cref{fig:discont}.

A final fix was applied in the $\phi$ dimension when extrapolating towards its bounds. Equally spaced bins ranging from \SIrange{0}{180}{\degree} are constructed along $\phi$, and receiver coordinates $\phi_k$ lie at the bin centers. The Mie model extrapolated beyond the first and last bin center to reach $\phi = 0^\circ$ and $180^\circ$. A simple solution to improve modeling near these boundaries was to reflect the nearest four data points across the boundary, thereby extending the support points in $\phi$ to beyond its domain, and perform the B-spline fit with those additional data points. This meant that the previous extrapolation to the bounds was replaced by interpolation between well-defined, tabulated values.

\begin{figure}[hbt]
\centering
\includegraphics[width=0.475\linewidth]{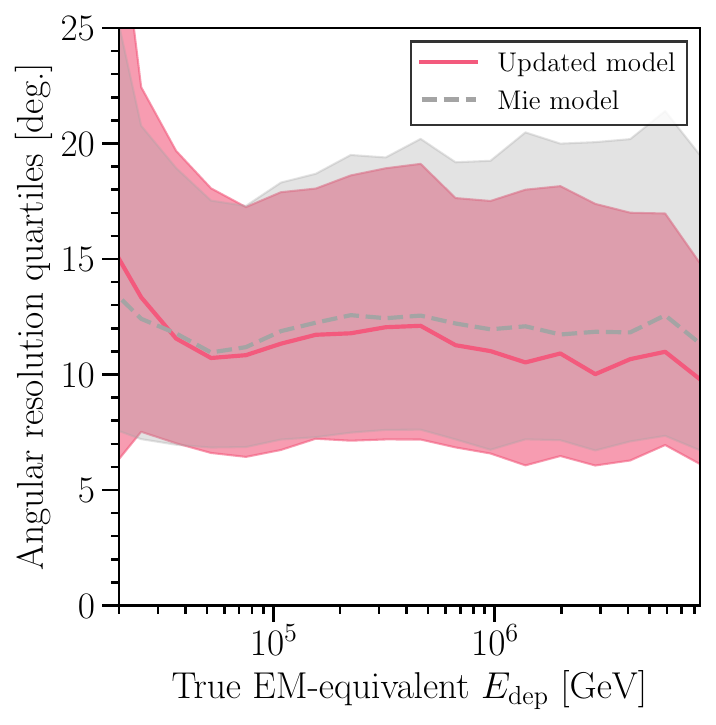}
\includegraphics[width=0.515\linewidth]{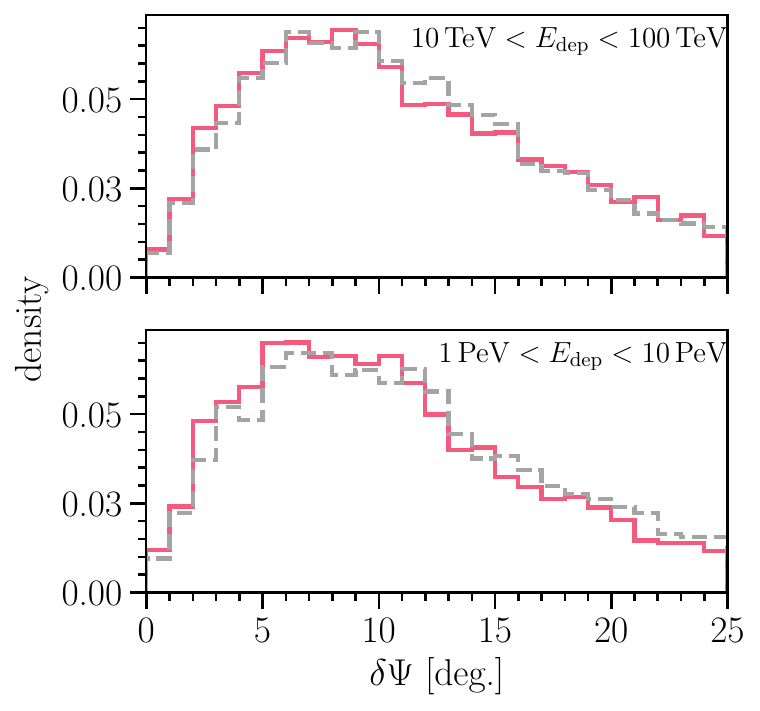}
\caption{The left panel shows quartiles of the distribution between reconstructed and true directions, $\delta \Psi$, as a function of deposited EM-equivalent energy for the benchmark MC sample described in the text. Results obtained using the Mie model are shown in gray. Results obtained with a newer ice model that includes the fixes described in \cref{sec:bspl} are shown in red. Neither include corrections to approximate the additional ice effects to be discussed in \cref{sec:bfr} and \cref{sec:tilt}. The solid red and dashed gray lines indicate their respective median angular resolutions. Lower (upper) cut offs for each color band show the respective 25 (75) percentile levels of the $\delta \Psi$ distribution. The right panels show distributions of $\delta \Psi$ in two different energy slices, between \SIrange{10}{100}{\tera \eV} (top) and \SIrange{1}{10}{\peta \eV} (bottom), with line colors and styles matching those of the left panel.}
\label{fig:bvmangres}
\end{figure}
\Cref{fig:bvmangres} illustrates the impact on the angular resolution going from the Mie model that is subject to interpolation and extrapolation artifacts (gray) to a newer ice model that includes the fixes described in this \lcnamecref{sec:bspl}. The left panel shows quartiles (25-50-75 percentile levels) of $\delta \Psi$ distributions for the benchmark MC as a function of $\edp$, obtained using the binning shown in the right panel of \cref{fig:recozenith}. The right panel shows distributions of $\delta \Psi$ in two different energy slices as described in the legend. We see that, in addition to the much improved zenith distributions shown in \cref{fig:recozenith}, the angular resolution is consistently worse with the Mie model, which exhibits additional degradation going towards higher energies.

\section{Approximation of ice anisotropies due to birefringence}
\label{sec:bfr}

Accurate characterization of physics quantities across a sparse array of PMTs requires accurate calibration of ice and instrument. This has been accomplished with calibration LEDs attached to a dedicated flasher board on each IceCube DOM~\cite{IceCube:2013llx} and by using large samples of downgoing minimum ionizing muons to set the global energy scale~\cite{Aartsen:2013vja}. The propagation of light in the glacial ice sheet is complex at Cherenkov wavelengths, with dependencies on depth that reflect Earth's climate across geological time scales as shown in \cref{fig:yield}. Absorption and scattering coefficients that describe the mean free path of a photon are fitted using calibration LED data over a range that encompasses all instrumented depths~\cite{IceCube:2013llx,tc-18-75-2024}. In general, the Cherenkov photon absorption (scattering) length in clear regions of the South Pole ice is longer (shorter) than in natural bodies of liquid water where neutrino detectors have been constructed or proposed~\cite{Spiering:2020iih}.

IceCube also discovered a directional dependence in the light propagation, or ice anisotropy, with maximal effect along the ice flow axis~\cite{Chirkin:2013lpu}. The flow axis is the direction that the bulk ice moves, which is at a rate of about \SI{10}{\m/yr}~\cite{doi:10.1126/science.1208336}. A microscopic explanation of the ice anisotropy due to birefringence of polycrystals was given in Ref.~\cite{Chirkin:2019vyq,tc-18-75-2024} and leads to improved agreement with calibration data over the previous phenomenological model~\cite{Chirkin:2013lpu}. However, as mentioned in \cref{sec:cha}, this breaks the azimuthal symmetry and naively would require another dimension, $\azi$, to be included in the model.

An alternative approach is to separate ice anisotropies from shower physics and take advantage of the symmetries inherent in both. The photoelectron yields from a particle shower can be evaluated in a bulk ice with no birefringence and flat, horizontal ice layers thus preserving azimuthal symmetry~\cite{Lundberg:2007mf}. The impact of ice birefringence can be estimated by simulating an isotropic, point-like light source and tabulating and fitting the corresponding photoelectron yields using two different ice models, one with birefringence and one without, including the improvements described in \cref{sec:extra}. An isotropic source simplifies the problem by removing the dependence on $\zen$. Comparisons of the two in amplitude, as done in Refs.~\cite{Usner:2018cel,Yuan:2022xdg}, and in shape, as discussed in \cref{sec:pdf}, allow for simple coordinate transformations that encode the impact of birefringence on Cherenkov photons. We emphasize that the key idea is the switch from a directional source to an isotropic source and that the birefringence corrections are computed using an isotropic source, which are later applied to directional sources. This is an approximation, but one which substantially improves the model and the angular resolution of showers, as will be shown.

\subsection{Amplitude}
\label{sec:amp}

The time-integrated photoelectron yield, or amplitude, at a receiver DOM differs depending on whether the ice is modeled with or without birefringence. Since the amplitude decreases monotonically with $r$, a translation to a larger (smaller) effective distance, $\reff$, can correct for too high (low) amplitude observed in the simplified ice model. Birefringence applies globally across the detector, independent of $(\xs,\ys)$~\cite{tc-18-75-2024}. Due to layer-by-layer differences in the ice, the $\zs$-dependence is kept. The goal, then, is to construct a function $\reff(\bvec{r}, \zs)$ that can be applied to correct amplitudes in the simplified model.
\begin{figure}[hbt]
\centering
\includegraphics[width=0.49\linewidth]{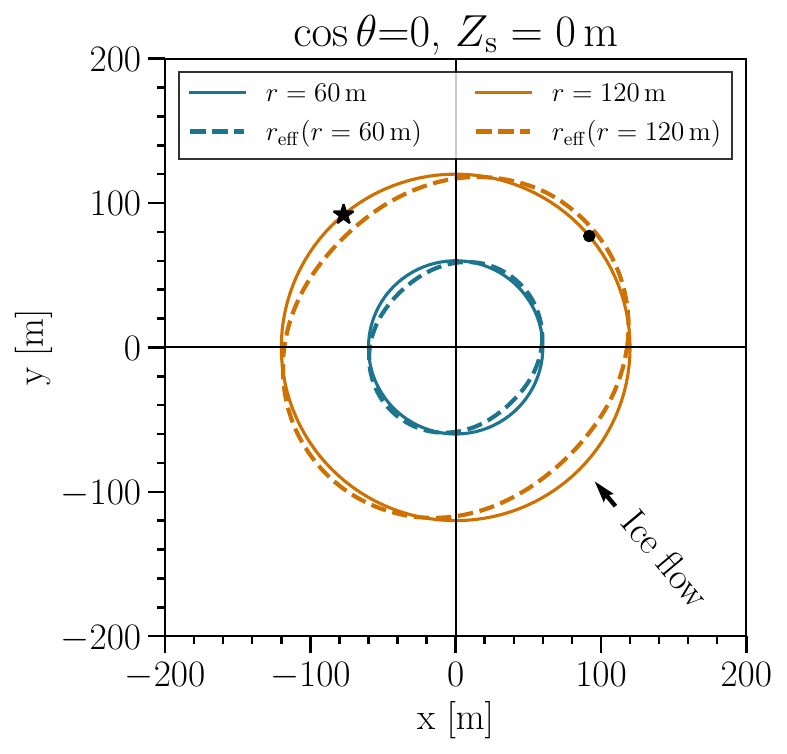}
\includegraphics[width=0.49\linewidth]{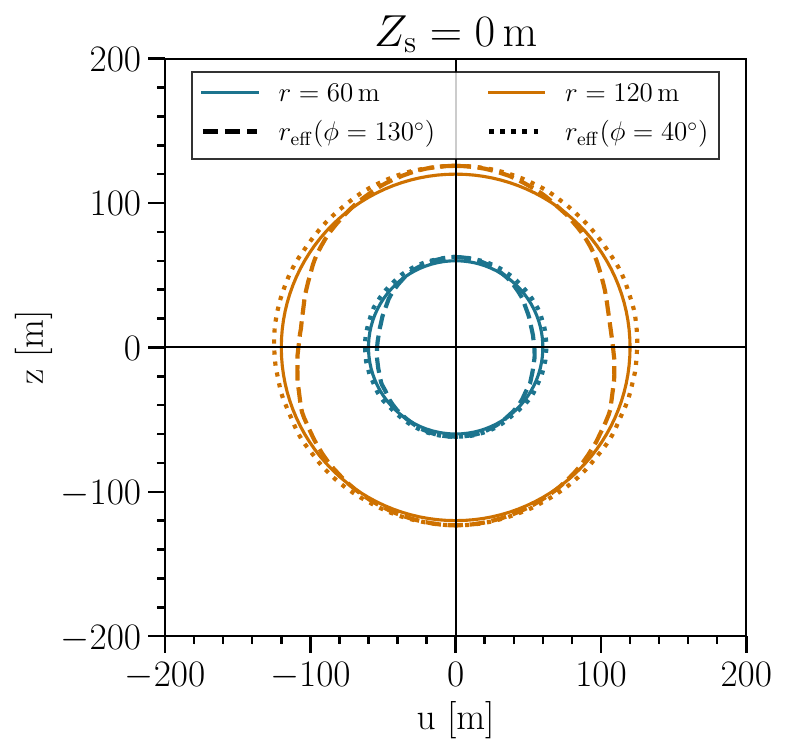}
\caption{A visualization of the effect of birefringence on $\reff$ for an isotropic light source centered at the origin of the detector. Solid lines are circles with radii of \SI{60}{\m} (blue) and \SI{120}{\m} (orange), centered at the origin. In the left panel, the dashed lines show the corresponding $\reff$ across the $xy$-plane. Notice that $\reff$ shrinks closer to the origin along the ice flow axis (black arrow), and shifts away from the origin perpendicular to the flow axis. The star (dot) indicates the location of the receiver in the left (right) panel of \cref{fig:tpdf}. In the right panel, a similar effect is observed along two planes perpendicular to the $xy$-plane, intersecting at the origin. The dashed (dotted) lines show $\reff$ in the $uz$-plane where $\hat{\bvec{u}} = \cos \phi \hat{\bvec{x}} + \sin \phi \hat{\bvec{y}}$ for $\phi = 130^\circ (40^\circ)$ denotes the unit vector parallel (perpendicular) to the ice flow axis shown in the left panel.}
\label{fig:reff}
\end{figure}

Two simulation sets are produced. Each consists of identical, isotropic, point-like light sources placed at a series of depths, $\zsj^*$, where $j$ is an index and the asterisk distinguishes this as referring to an isotropic source. One set assumes the simplified ice model, which does not include birefringence or layer undulations. The other includes birefringence but no layer undulations (see \cref{sec:tilt}). The time-integrated amplitudes are tabulated at receivers placed in spherical coordinates $\bvec{r}_k = (r_k, \theta_k, \phi_k)$ from the isotropic source, with the polar angle $\theta$ now defined relative to $\hat{\bvec{z}}$. A tensor-product B-spline is then fit to the tabulated data in order to construct a smooth approximator for the isotropic source amplitudes, $N_{\text{bfr}}(\bvec{r}, \zsj^*)$ and $N_\text{simple}(\bvec{r}, \zsj^*)$, with and without birefringence respectively. At each simulated $\zsj^*$, these amplitudes are compared across the two ice models to solve for $\reff(\bvec{r}_k, \zs^*)$ such that
\begin{equation}
    N_{\text{simple}}\Bigl(\reff \bigl(\bvec{r}_k, \zsj^* \bigr), \theta_k, \phi_k, \zsj^*\Bigr) = N_{\text{bfr}}\Bigl(\bvec{r}_k, \zsj^* \Bigr).
    \label{eq:reff}
\end{equation}
Once evaluated across the grid of receiver positions, the discrete set of $\reff(\bvec{r}_k, \zsj^*)$ are fitted again with tensor-product B-splines for each $\zsj^*$ and then stacked across $\zsj^*$ to give $\reff(\bvec{r}, \zs^*)$.

The obtained $\reff$ is presented for a slice across $\cos \theta=0$ and $\zs^*=\SI{0}{\m}$ (center of IceCube) in the left panel of \cref{fig:reff}. Two circles are drawn as solid lines at $r=\SI{60}{\m}$ (blue) and \SI{120}{\m} (orange). The corresponding $\reff$ at those distances are shown as dashed lines in blue and orange, respectively. The direction of ice flow is indicated by the arrow. The right panel includes a similar visualization, only now in two planes that lie perpendicular to the horizontal plane shown in the left panel. The dashed (dotted) line shows $\reff$ in the vertical plane that lies along (perpendicular to) the flow axis, at $\phi=130^\circ \: (40^\circ)$~\cite{tc-18-75-2024}. Thus, the right panel displays two $uz$-planes, where $u$ is the distance along the direction given by $\hat{\bvec{u}} = \cos \phi \hat{\bvec{x}} + \sin \phi \hat{\bvec{y}}$. The effect of birefringence is to shrink the distance along the ice flow axis, leading to a larger amplitude, while slightly increasing the distance perpendicular to the flow axis, leading to a smaller amplitude. These effects are consistent with observations from calibration data~\cite{tc-18-75-2024}.

\subsection{Time probability density function}
\label{sec:pdf}

The $\reff$ discussed in \cref{sec:amp} corrects for amplitude differences due to birefringence but does not account for differences in the time profile. These photon arrival time distributions, i.e.\ the probability density function (PDF) in $\tres$, add timing information to the model (see \cref{eq:bulk}).
\begin{figure}[hbt]
\centering
\includegraphics[width=0.49\linewidth]{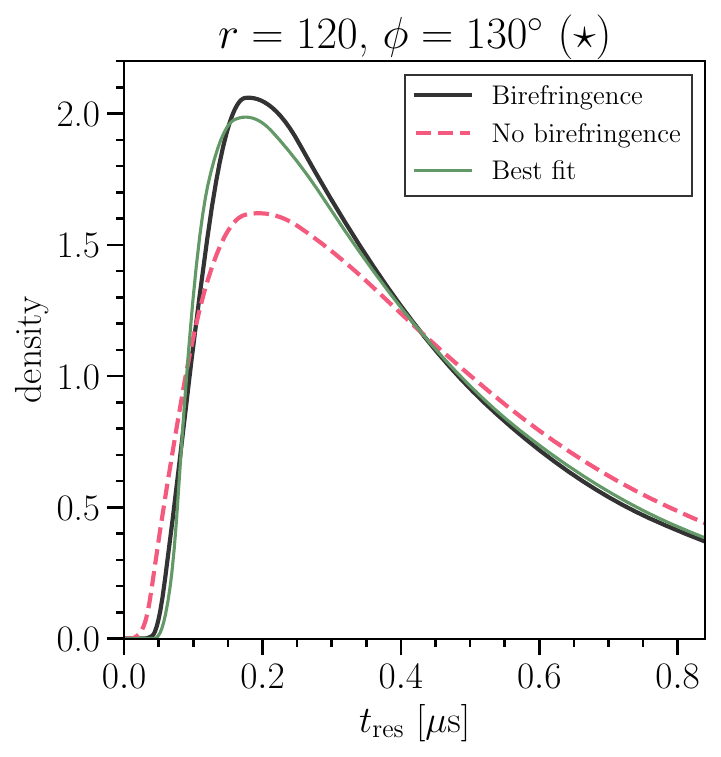}
\includegraphics[width=0.49\linewidth]{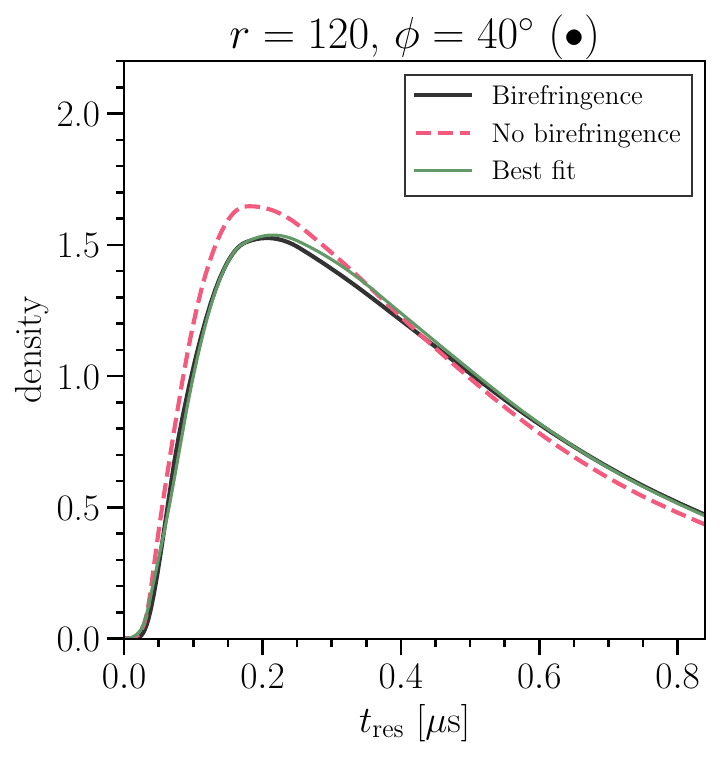}
\caption{Impact of birefringence on the arrival time PDF from an isotropic point-like light source, at a source depth of $\zs=0$ and the receiver coordinates indicated in the titles. For reference, the location of the receiver in the left (right) panel is indicated by the star (dot) in the left panel of \cref{fig:reff}. The black, solid (red, dashed) line shows the time PDF with (without) birefringence simulated. The green, solid line shows the result of a two-parameter translation in $r$ and $t$, which when applied to the time PDF without birefringence gives the best agreement with the black line.}
\label{fig:tpdf}
\end{figure}

The impact of birefringence on the PDF is shown in \cref{fig:tpdf} for two azimuthal directions as given in the titles. Their relative positions to the source are indicated by the star and dot in the left panel of \cref{fig:reff}. The impact of birefringence on the time PDF is illustrated going from the red, dashed to the black, solid lines, with a larger difference visible in the left panel, when the source-receiver direction lies along the flow axis. Notably, along the ice flow axis, the time PDF is squeezed narrower while preserving the mode of the distribution. It is not possible to capture such an effect with a sole translation in $r$. In order to obtain a narrower distribution, a smaller source-receiver distance is needed, which in turn causes a shift in the mode.

A natural generalization to model shape differences is to introduce a translation in $t$ in addition to $r$. That is, instead of a single-parameter correction $\reff(\bvec{r}, \zs^*)$ we seek a two-parameter correction $\rmod(\bvec{r}, \zs^*)$ and $\tmod(\bvec{r}, \zs^*)$, which can be applied simultaneously to better approximate the $\tres$ PDF. Using the same two simulation sets described in \cref{sec:impact}, and extending the construction to include timing information, we generate CDFs $G_\text{bfr}(\bvec{r}, \tres, \zsj^*)$ and $G_\text{simple}(\bvec{r}, \tres, \zsj^*)$. The corrections are obtained by Nelder-Mead~\cite{gao2012implementing} minimization over the Kolmogorov–Smirnov (KS) statistic for receivers on $\bvec{r}_k$. Specifically, for each simulated $\zsj^*$, we find $\rmod(\bvec{r}_k, \zsj^*)$ and $\tmod(\bvec{r}_k, \zsj^*)$ that minimizes
\begin{equation}
    \sup_{\tres \in [0, 1.5 T]}\Bigl|G_\mathrm{simple}\Bigl(\rmod \bigl(\bvec{r}_k, \zsj^* \bigr), \theta_k, \phi_k, \tres-\tmod \bigl(\bvec{r}_k, \zsj^*\bigr), \zsj^*\Bigr) - G_\mathrm{bfr}\Bigl(\bvec{r}_k, \tres, \zsj^*\Bigr)\Bigr|,
    \label{eq:ks}
\end{equation}
where $T = \operatorname*{arg\,max}_{\tres} g_\text{bfr}(\bvec{r}_k, \tres, \zsj^*)$ is the mode of the $\tres$ PDF, which is denoted here as $g \equiv \partial G/\partial \tres$. Restricting the range over which the KS statistic is computed to $[0, 1.5 T]$ allows the fit to optimally describe the PDF over the region where most of the density is concentrated. The result is illustrated by the green lines in \cref{fig:tpdf}, which show $g_\text{simple}(\rmod, \tres-\tmod, \zsj^*=0)$ at the best-fit values for $\rmod$ and $\tmod$ obtained by minimizing \cref{eq:ks}, for the receiver coordinates indicated in the panel titles. While slight differences still exist compared to the $\tres$ PDFs with birefringence (black lines), the agreement is significantly improved over the raw $\tres$ PDF without birefringence (red, dashed lines). For each $\zsj^*$, the obtained discrete sets of $\rmod(\bvec{r}_k, \zsj^*)$ and $\tmod(\bvec{r}_k, \zsj^*)$ are separately fitted across receivers $k$ with tensor-product B-splines. Finally, these surfaces are stacked across $\zsj^*$ to give $\rmod(\bvec{r}, \zs^*)$ and $\tmod(\bvec{r}, \zs^*)$.
\begin{figure}[hbt]
\centering
\includegraphics[width=0.475\linewidth]{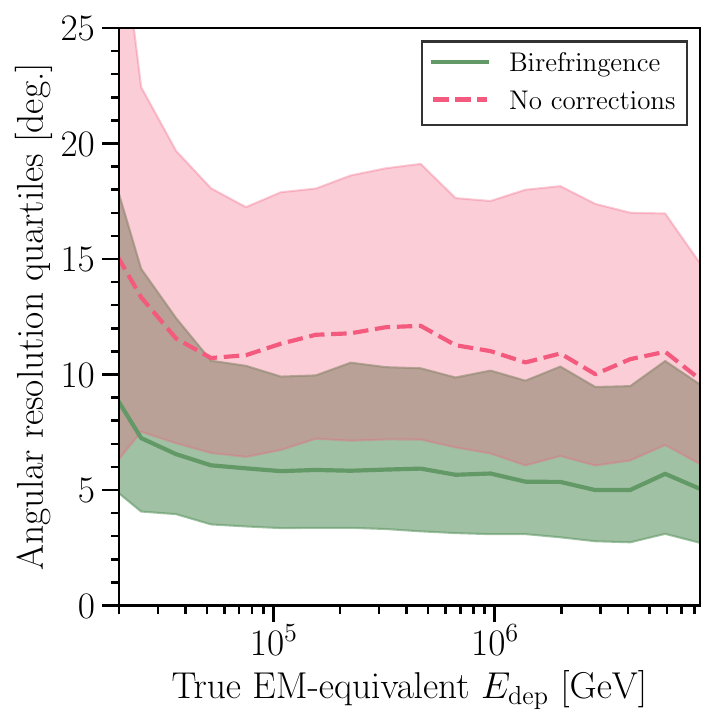}
\includegraphics[width=0.515\linewidth]{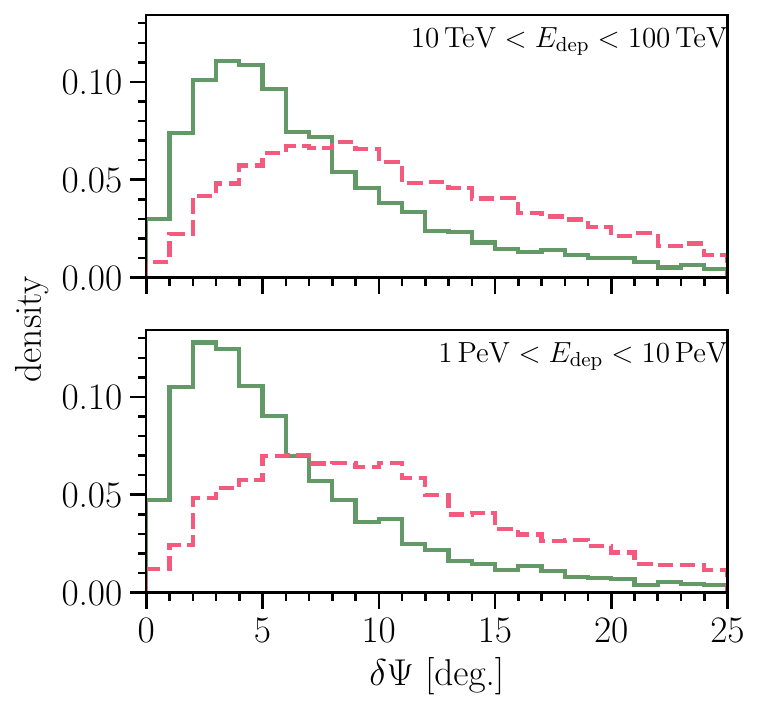}
\caption{Same as \cref{fig:bvmangres} but now comparing the angular resolution using \cref{eq:flat} (green, solid) to \cref{eq:bulk} (red, dashed), where both include the fixes described in \cref{sec:bspl}. In the left panel, a substantial improvement is observed when birefringence corrections are included. Note the red intervals are identical to those in \cref{fig:bvmangres}. The right panels show distributions of $\delta \Psi$ in two different energy slices, between \SIrange{10}{100}{\tera \eV} (top) and \SIrange{1}{10}{\peta \eV} (bottom), with line colors and styles matching those of the left panel.}
\label{fig:fvbangres}
\end{figure}

The functions $\reff(\bvec{r}, \zs^*)$, $\rmod(\bvec{r}, \zs^*)$ and $\tmod(\bvec{r}, \zs^*)$ can be jointly used to approximate the effects of ice birefringence on shower photoelectron yields and time profiles. In order to substitute $\zs^* \rightarrow \zs$, we need to assume that the corrections obtained with an isotropic source generalize to a directional source. To justify this assumption, note that these transformations are translations, rather than rotations. A modification of the radial and time coordinates passed to the amplitude and CDF given in \cref{eq:bulk} does not destroy information about the directionality of Cherenkov photon emission from particle showers. Thus, \cref{eq:bulk} becomes
\begin{equation}\label{eq:flat}
\eqflat{\zs}.
\end{equation}
Here we have used the notation $\bvec{r}_z$ to refer to the source-receiver vector in the standard spherical coordinate system with polar axis $\hat{\bvec{z}}$, in contrast to $(\theta, \phi)$, which are defined in the spherical coordinate system with polar axis $\hat{\bvec{p}}$ as shown in \cref{fig:coordinates}. Finally, since $\reff$, $\rmod$ and $\tmod$ are differentiable functions, the Jacobian can be computed for gradient-based minimization.

To validate that the corrections described in this \lcnamecref{sec:bfr} lead to an improved description of particle showers in IceCube, we compared the directional reconstruction performance of the benchmark MC set introduced in \cref{sec:impact}. \Cref{fig:fvbangres} shows the angular resolution, $\delta \Psi$, with (green, solid) and without (red, dashed) birefringence correction. The left panel shows the median angular resolution (solid and dashed lines) as a function of $\edp$. The bands correspond to the \SI{25}{\%} and \SI{75}{\%} intervals in the $\delta \Psi$ distribution. The right panels show the full $\delta \Psi$ distribution in two energy regimes, between \SIrange{10}{100}{\tera \eV} (top) and \SIrange{1}{10}{\peta \eV} (bottom), with line colors and styles matching those of the left panel. From these comparisons, it is evident that a substantial improvement in the directional reconstruction is achieved when using \cref{eq:flat} over \cref{eq:bulk}, one which far exceeds the improvement from additional photon statistics going from lower to higher energies.

\section{Approximation of ice layer undulations}
\label{sec:tilt}

Polar ice stratigraphy exhibits a strong dependence on depth, broadly reflecting changes in Earth's climate over time. Layers of relatively homogeneous ice isochrons---ice layers that formed at around the same time period---are each modeled with unique scattering and absorption coefficients. The elevation change for each layer was initially calibrated using data from a laser dust logger deployed in seven IceCube boreholes~\cite{IceCube:2013jrb}. More recently, a reevaluation of ice optical properties with in situ calibration data revealed that ice layers exhibit undulations, believed to be formed by geological structures in the bedrock, across the detector array rather than solely along a single axis~\cite{IceCube:2023qua}. As layer-by-layer ice property differences can result in significant differences in the detected photoelectron yields (see \cref{fig:yield}), an accurate description is important for simulation and reconstruction.

\subsection{Parameterization of the ice depth dependence}
\label{sec:pref}

\begin{figure}[hbt]
\centering
\includegraphics[width=0.49\linewidth]{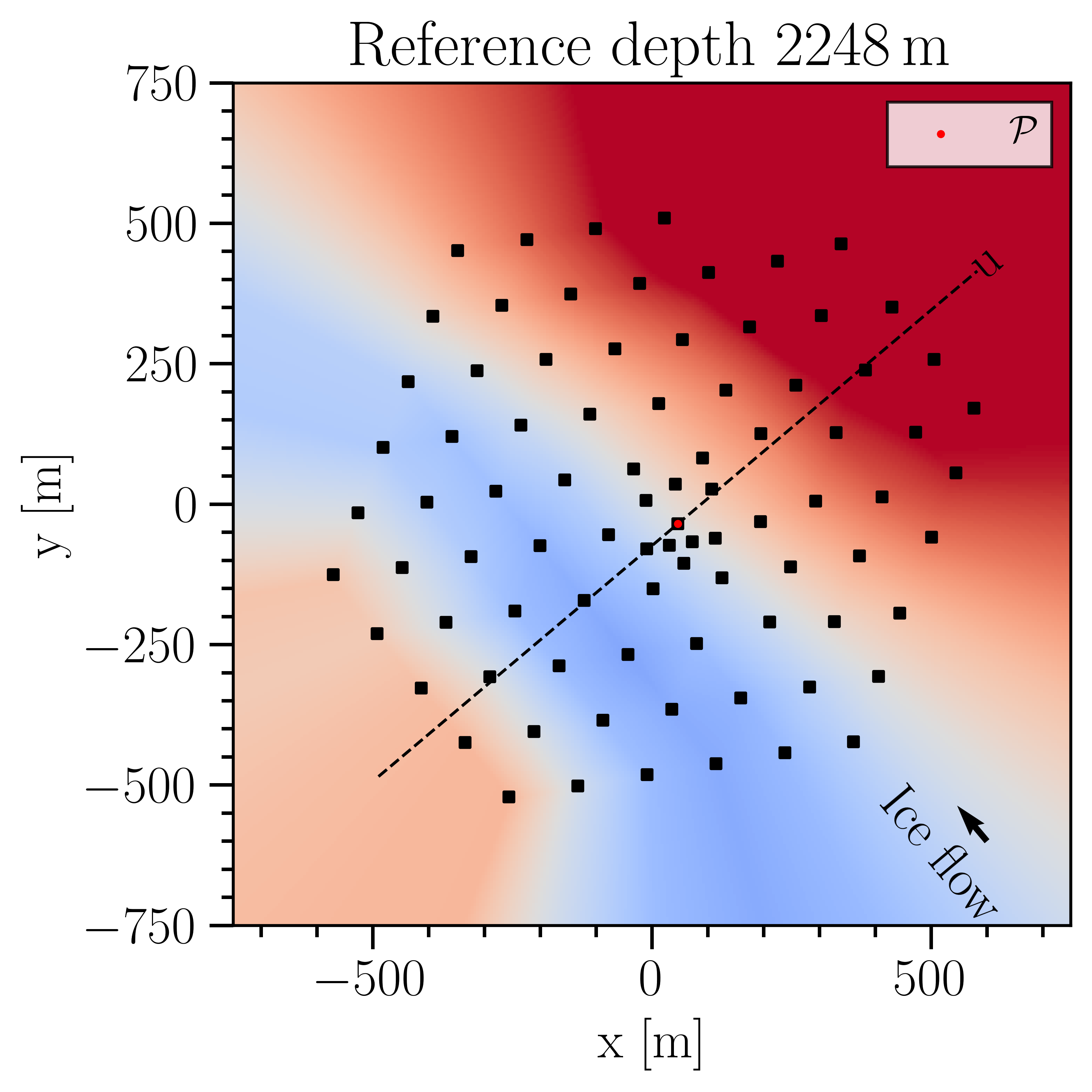}
\hspace{-0.2cm}
\includegraphics[width=0.42\linewidth,angle=90,origin=ll,trim=0 0.8cm 0 0]{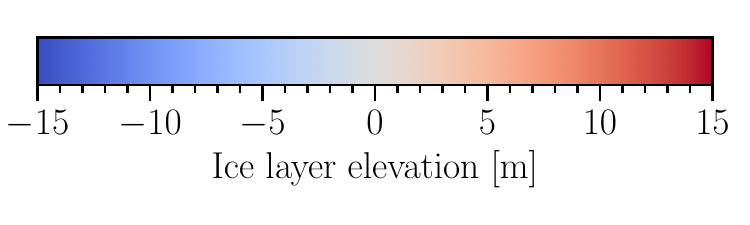}
\hspace{0.2cm}
\includegraphics[height=0.47\linewidth]{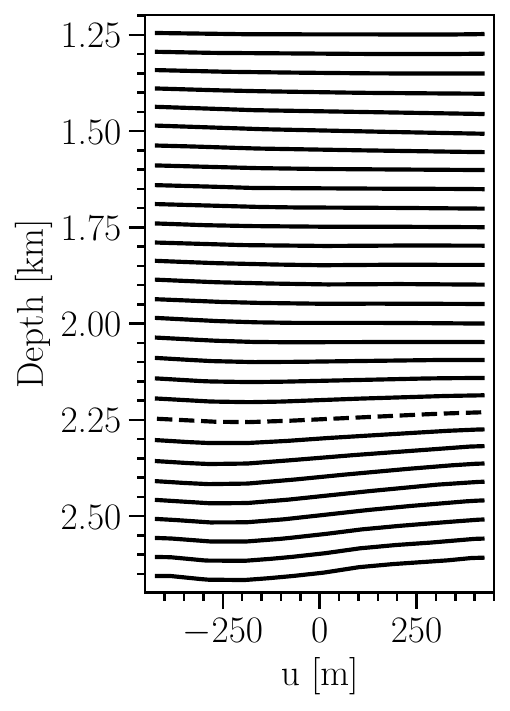}
\caption{The left panel shows the elevation change for an ice isochron at a $\Pref$-depth of \SI{2248}{\m}, where $\Pref$ is the reference point indicated by the red dot. Black squares corresponds to the $(x,y)$ positions of IceCube strings, and the color map shows the elevation change relative to $\zref$. The region outside the instrumented footprint is extrapolated~\cite{IceCube:2023qua}. The dashed line indicates the $u$-axis shown in the right panel, which lies perpendicular to the ice flow direction. The right panel shows depths of ice isochrons along $\hat{\bvec{u}} = \cos \phi \hat{\bvec{x}} + \sin \phi \hat{\bvec{y}}$, where $\phi=40^\circ$, and the dashed line corresponds to the layer shown in the left panel. The visualized plane is perpendicular to the ice flow axis and intersects $\Pref$ at $u=\SI{0}{\m}$. Note that the relative change in each layer is stronger at the deeper regions of the instrumented volume.}
\label{fig:tilt}
\end{figure}

Since ice isochrons are not perfectly flat, optical properties of the ice are defined as a function of depth (or $z$ in detector coordinates) at a fixed reference point, $\Pref$, near the origin in the $xy$-plane and shown as the red dot in the left panel of \cref{fig:tilt}~\cite{IceCube:2013llx,IceCube:2023qua}. In other words, the ice model is parameterized as a function of what we will refer to as the $\Pref$-depth. A mapping of ice layer undulations is needed to convert a physical position, $(x,y,z)$, back to the appropriate $\Pref$-depth, $\zref(x, y, z)$, to retrieve the corresponding ice optical properties at that position. Such a mapping can then be used to apply the correct scattering and absorption lengths for any position throughout the detector. 

The left panel of \cref{fig:tilt} shows the ice layer elevation change across the $xy$-plane for the isochron at a $\Pref$-depth of \SI{2248}{\m} ($\zref=\SI{-300}{\m}$ in detector coordinates). Specifically, the color map shows $\zice - \zref(x, y, \zice)$, where $\zice$ is defined such that $\zref(x, y, \zice)=\SI{-300}{\m}$. Black squares indicate the location of the 86 strings and the color map shows the ice layer elevation change at different $(x,y)$ locations. The black arrow indicates the direction of ice flow, and the dashed line indicates the $u$-axis shown in the right panel. The right panel of \cref{fig:tilt} shows the depths of ice isochrons in the plane normal to the flow direction and intersecting $\Pref$ at $u=\SI{0}{\m}$, where $\hat{\bvec{u}} = \cos \phi \hat{\bvec{x}} + \sin \phi \hat{\bvec{y}}$ and $\phi = 40^\circ$. Note that the relative change in each layer increases towards at the deeper regions of the detector.

Within the hexagonal footprint of the instrumented region $\zref$ is linearly interpolated across a grid of equilateral triangles roughly coinciding with the string locations, while outside the instrumented region $\zref$ is extrapolated by expanding the detector's hexagonal outline radially and matching to the corresponding value on the hexagon boundary~\cite{IceCube:2023qua}. Such an extrapolation ensures well-defined behavior everywhere. A mismatch between the extrapolated model and reality could affect the reconstruction of events outside the detector. For events contained within a fiducial volume, as is the case in the benchmark MC used throughout this work, the impact of extrapolation is expected to be negligible.

\subsection{Correcting for undulating ice layers}
\begin{figure}[hbt]
\centering
\includegraphics[width=0.49\linewidth]{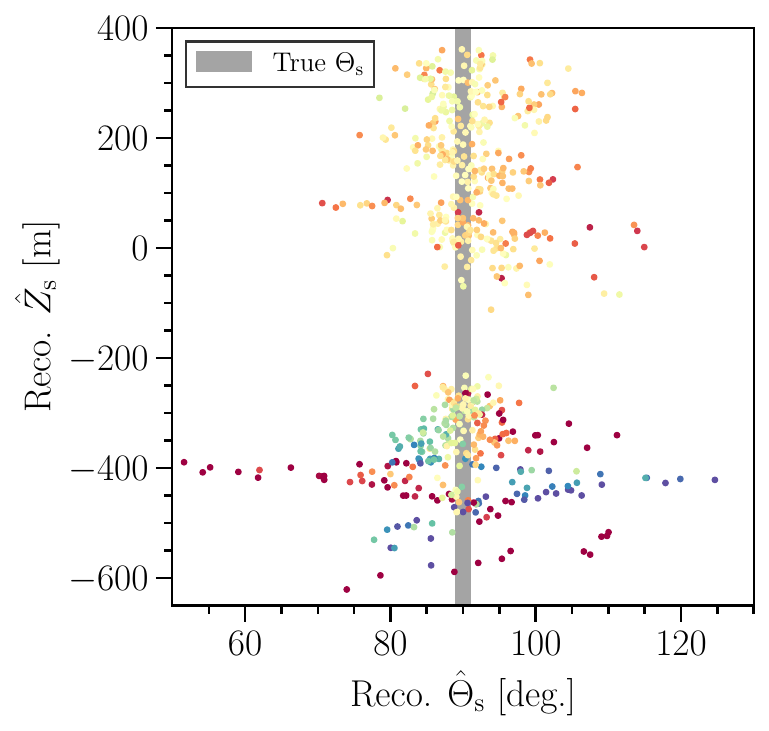}
\includegraphics[width=0.49\linewidth]{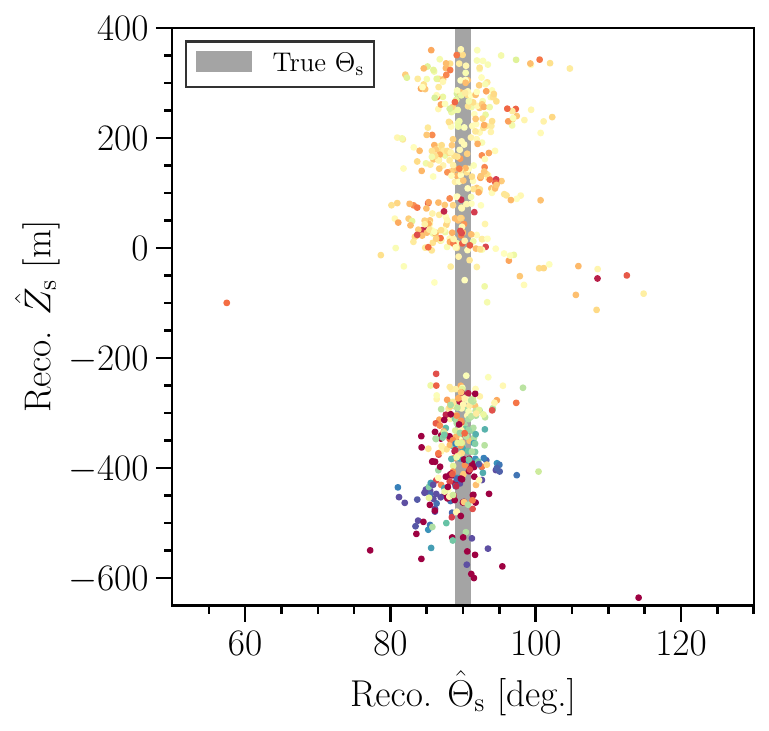}

\includegraphics[width=0.49\linewidth]{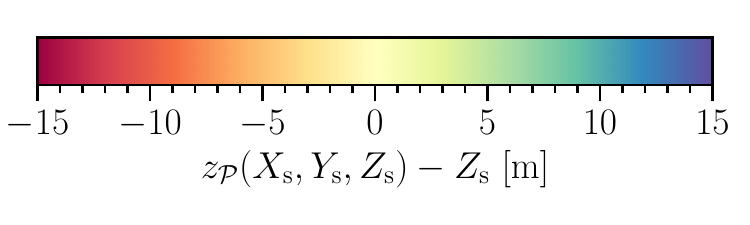}
\caption{Scatter plots of reconstructed $\hatzen$ and $\hatzs$ for a subset of the benchmark MC, defined to be those with true $|\zen - 90^\circ| < 1^\circ$ as indicated by the vertical shaded region. The gap around $\hatzs=-150$ is due to a region of heightened scattering and absorption stretching horizontally across the detector that is excluded from the selection~\cite{IceCube:2020wum}. The left panel uses \cref{eq:flat}, while the right panel uses \cref{eq:full}. The variance in $\hatzen$ is larger in the left panel and shows a clear correlation with $\zref(\xs, \ys, \zs) - \zs$, which is calculated based on each event's true position and indicated by the color. Pulls on $\hatzen$ are reduced when using \cref{eq:full}, as shown in the right panel. The depth dependence in the left panel is attributable to stronger layer undulations towards deeper regions of the detector, as shown in the right panel of \cref{fig:tilt}.}
\label{fig:rzenrzknot}
\end{figure}
In MC simulations, as a photon propagates through the ice, its physical position in the detector at every step is passed to $\zref$ to look up which ice layer it belongs to at $\Pref$. Fast event reconstruction routines do not track photons individually, but can approximate the effect by using the source position instead to obtain $\zref(\xs, \ys, \zs) = \zref(\bvecXs)$. Substituting into \cref{eq:flat} we obtain
\begin{equation}\label{eq:full}
\eqflat{\zref(\bvecXs)},
\end{equation}
which gives the functional form of the photoelectron yield correcting for both birefringence and ice layer undulations. As $\zref$ is a function constructed by linear interpolation between fixed points in position space, it is not smooth everywhere~\cite{IceCube:2023qua}. In practice, this does not pose an issue for computing gradients, since discontinuities in its derivatives are small and only rarely occur. Thus, we can compute the Jacobian terms $\partial \zref / \partial X_i$, where $X_i \in \{\xs, \ys, \zs\}$, for use in gradient-based minimizers.

Factorizing out ice layer undulations from the B-spline model itself makes it extremely simple to switch to newer $\zref(x, y, z)$ models as well. If, instead, one were to generate photon-yield expectations without this factorization, then for each update to $\zref(x, y, z)$ a massive resimulation campaign would be needed to generate the raw data to construct an updated model. The dimensionality would also need to be increased to include $(\xs, \ys)$. The cost of the factorization is that using only the source position of the cascade is an approximation, as the path between source and receiver DOM can traverse several layers of ice with varying amounts of undulations. This can be seen in the right panel of \cref{fig:tilt}. Fortunately, the ice layer undulations are gradual and this approximation improves for shorter source-receiver distances, which is also where most of the statistics used in event reconstruction are expected. 

\begin{figure}[hbt]
\centering
\includegraphics[width=0.475\linewidth]{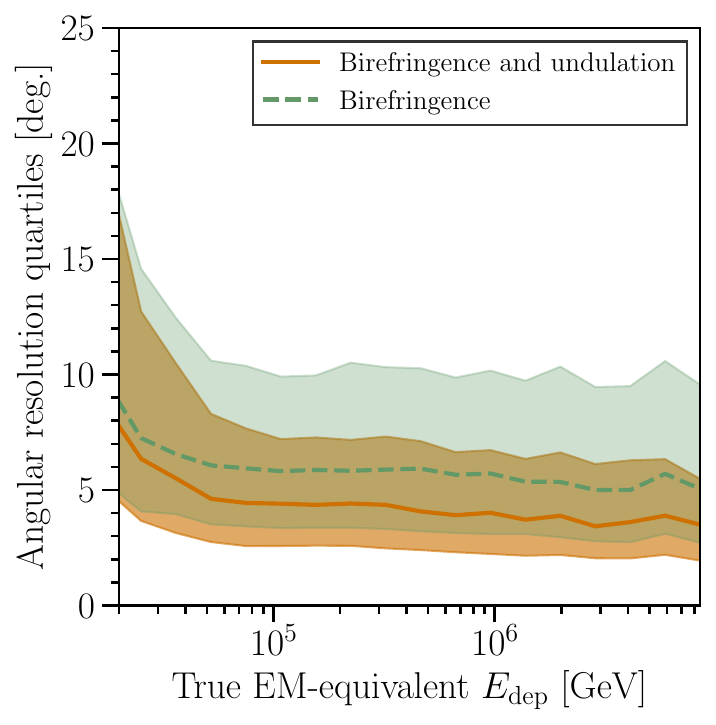}
\includegraphics[width=0.515\linewidth]{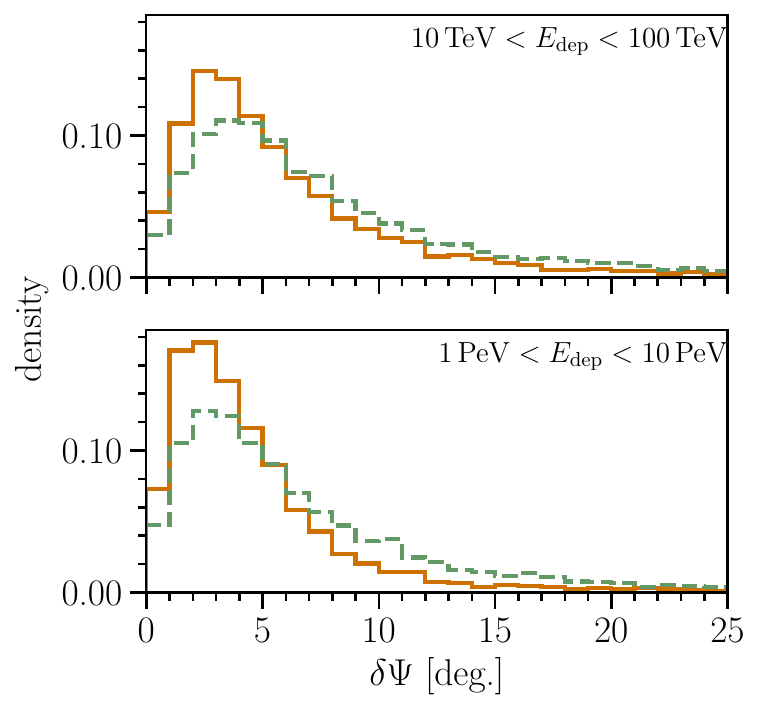}
\caption{Same as \cref{fig:fvbangres} but now comparing the angular resolution using \cref{eq:full} (orange, solid) to \cref{eq:flat} (green, dashed). In the left panel, an improvement is observed in the $\delta \Psi$ distribution quartiles when ice layer undulations are additionally included. Note the green intervals are identical to those in \cref{fig:fvbangres}. The right panels show distributions of $\delta \Psi$ in two different energy slices, between \SIrange{10}{100}{\tera \eV} (top) and \SIrange{1}{10}{\peta \eV} (bottom), with line colors and styles matching those of the left panel.}
\label{fig:fvfangres}
\end{figure}
To validate \cref{eq:full}, we again use the same benchmark MC as described in \cref{sec:impact}. The impact of ice layers is easily seen in reconstructed zenith, $\hatzen$, as a function of $\hatzs$. \Cref{fig:rzenrzknot} shows the improvement in $\hatzen$ across $\hatzs$ going from \cref{eq:flat} (left panel) to \cref{eq:full} (right panel) for a subset of events with true $|\zen - 90^\circ| < 1^\circ$. Each point corresponds to an event's reconstructed quantities, and its color indicates $\zref(\xs, \ys, \zs) - \zs$. In the left panel, a larger variance in $\hatzen$ shows a visible correlation with $\zref(\xs, \ys, \zs) - \zs$. When this is accounted for by the correction in \cref{eq:full}, a much improved clustering of $\hatzen$ near true $\zen$ is obtained, as shown in the right panel. The outlier visible in the right panel at $\hatzs \approx \SI{-100}{\m}$ can be attributed to a local minimum, which can exist in the likelihood surfaces used during event reconstruction. Methods to reduce the possibility of getting stuck in local minima include running additional iterations of the minimizer or improving the seed that is used to initialize the minimizer.

Similar to \cref{fig:fvbangres}, the improvement in angular resolution for the full benchmark MC is shown in \cref{fig:fvfangres} as a function of $\edp$ (left panel) and in two different energy regimes (right panels). Distributions of $\delta \Psi$ with (orange, solid) and without (green, dashed) ice layer undulation correction are compared. The left panel shows the median angular resolution (solid and dashed lines) as a function of $\edp$. The bands correspond to the \SI{25}{\%} and \SI{75}{\%} intervals in the $\delta \Psi$ distribution. The right panels show the $\delta \Psi$ distribution in two energy regimes, between \SIrange{10}{100}{\tera \eV} (top) and \SIrange{1}{10}{\peta \eV} (bottom), with line colors and styles matching those of the left panel. From these comparisons, it is evident that including corrections to account for ice layer undulations, as in \cref{eq:full}, leads to a further improvement in the directional reconstruction over \cref{eq:flat}.

\section{Further considerations}
\label{sec:considerata}
\subsection{Shower longitudinal extension}
\label{sec:dc}

Until now, the discussion has focused on applying a single, point-like cascade in the reconstruction of particle showers. By including approximations of the ice birefringence and ice layer undulations, we see a substantial improvement in the angular resolution and likelihood description, as shown in \cref{fig:iota}. In reality, particle showers have energy-dependent longitudinal and transverse extensions. At high energies, the extension is predominantly longitudinal, along the primary momentum direction~\cite{Radel:2012ij}. While this length is typically much smaller than the IceCube string spacing, it does provide some additional information for event reconstruction.

\begin{figure}[hbt]
\centering
\includegraphics[width=0.49\linewidth]{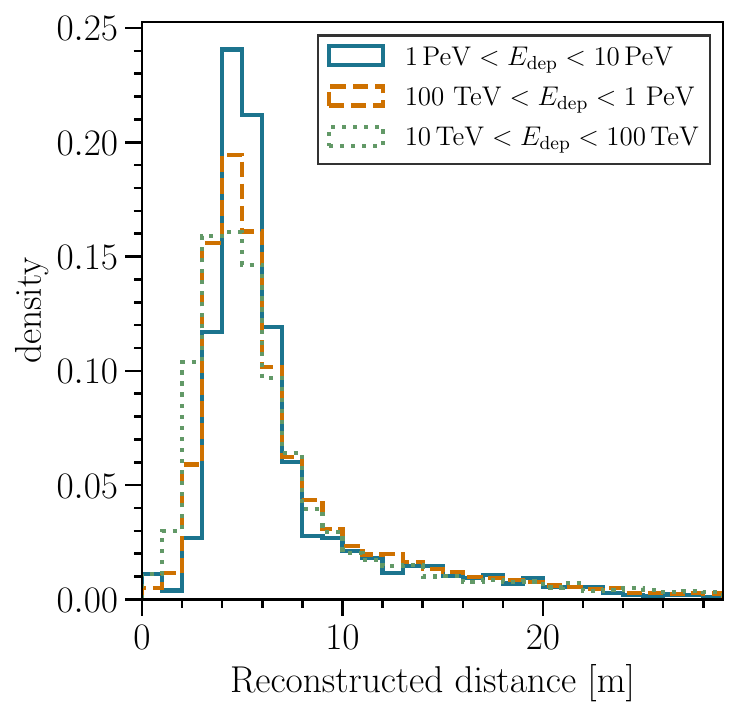}
\includegraphics[width=0.49\linewidth]{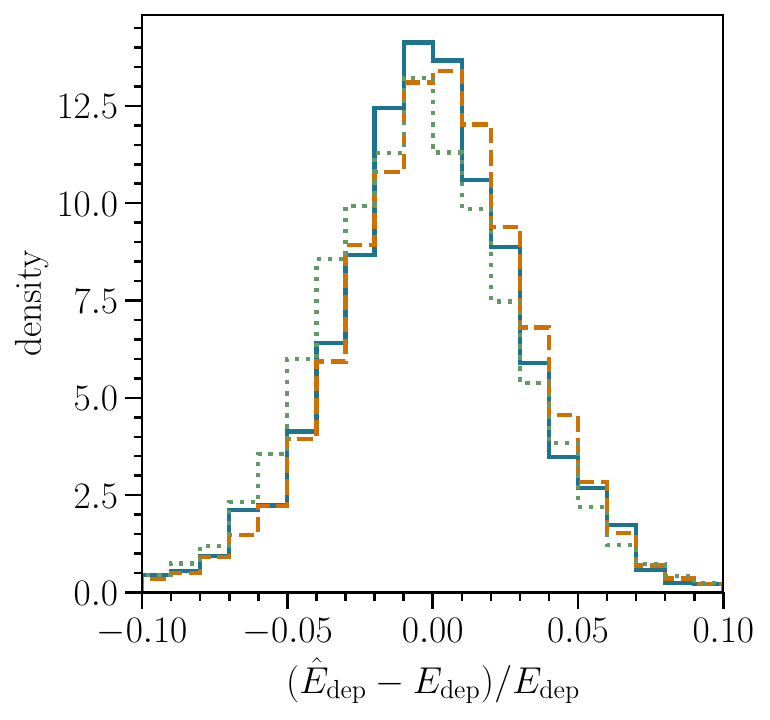}
\caption{Using the benchmark MC, the left panel shows the distribution of the reconstructed distance between two cascades for three different energy regimes. There is a trend towards slightly longer distances as the true EM-equivalent deposited energy, $\edp$, increases, which follows expectations from shower physics~\cite{Radel:2012ij}. The right panel shows distributions of the pulls on reconstructed energy, denoted as $\hatedp$, for the same energy slices as in the left panel, illustrating that its resolution is on the order of a few percent.}
\label{fig:dcslices}
\end{figure}
One rationale for the point-like cascade approach is simplicity; it allows for the photoelectron yields to scale linearly with energy~\cite{Aartsen:2013vja}. Encoding an energy-dependent shower extension into the model would introduce more complexity. Furthermore, due to shower stochastics that can cause per-event fluctuations in the shower profile, such an encoding may not be broadly applicable to data without additional modifications to account for the stochasticity.

Here we employ a simple two-cascade model to approximate the longitudinal extension of particle showers. Originally this was developed for tau neutrino reconstruction, wherein a $\nu_\tau$ CC interaction can produce two distinct particle showers, the first at the interaction vertex and the second upon the subsequent tau decay, separated by the travel distance of the tau lepton~\cite{hallen2013measurement}. This is modeled as two cascades with the same direction separated by a variable distance. The energies of the cascades and their separation distance are fitted as free parameters. Naturally, this approach should be able to model generic longitudinal extensions in particle showers, and indeed we see some improvement in the angular resolution when applied to the $\nu_e$ benchmark MC.

The two-cascade model relies on \cref{eq:full} for the description of both cascades. We first perform a single-cascade reconstruction with all ice-associated corrections, then use that as a seed for the two-cascade routine. As a final step, the likelihood of both single- and two-cascade reconstructions are compared and best fit is chosen. In the vast majority of cases, the two-cascade routine is preferred. The benchmark MC introduced in \cref{sec:impact}, which, like all standard IceCube MC, includes a simulation of the average longitudinal profile of particle showers~\cite{Radel:2012ij} but not their shower-to-shower fluctuations aside from those due to the Landau-Pomeranchuk-Migdal effect~\cite{Klein:1998du,Workman:2022ynf} for EM showers with energies above \SI{1}{\peta \eV}, is used for validation. \Cref{fig:dcslices} shows cross checks of the reconstructed separation distance distributions (left panel) and pulls on $\edp$, the EM-equivalent deposited energy (right panel). In both panels, three different $\edp$ slices are shown. The reconstructed distance distribution is pulled to larger values with increasing $\edp$, though all peak below \SI{10}{\m}, consistent with expectations from $\nu_e$ simulations.

\begin{figure}[hbt]
\centering
\includegraphics[width=0.475\linewidth]{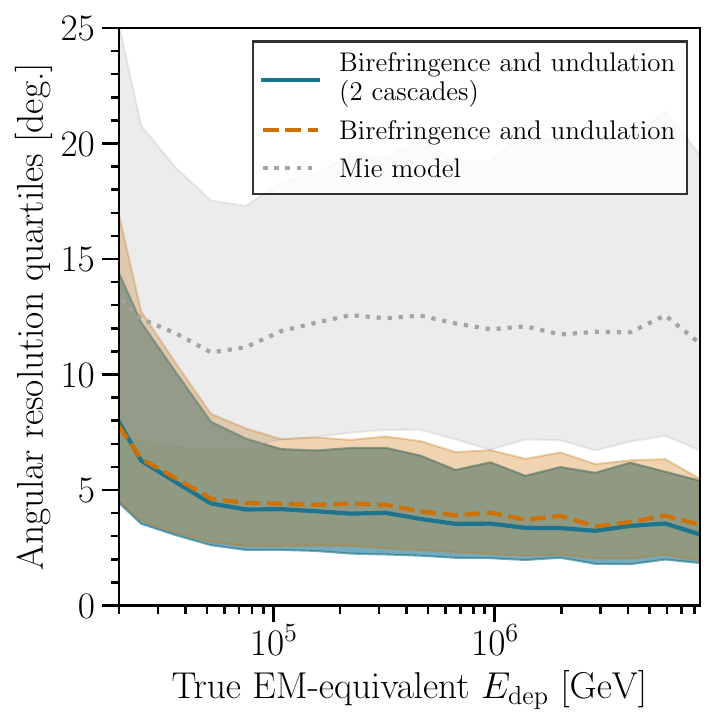}
\includegraphics[width=0.515\linewidth]{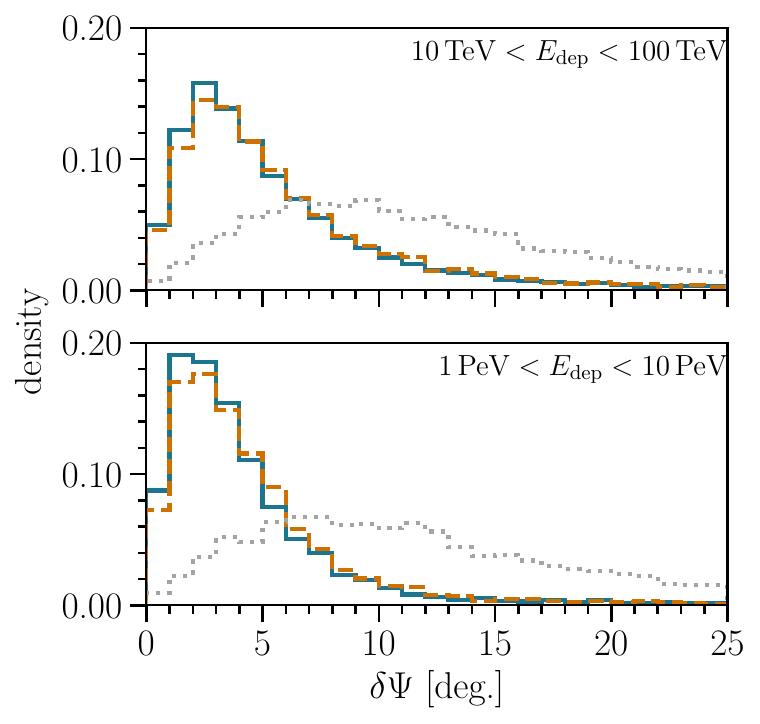}
\caption{Same as \cref{fig:fvfangres} but now comparing the angular resolution using \cref{eq:full} with a two-cascade model (blue, solid) to a single cascade model (orange, dashed). For reference, results obtained with the Mie model without corrections are shown in dotted gray. The left panel shows that the most accurate $\delta \Psi$ quartiles are obtained with a two-cascade model. Note the orange intervals are identical to those in \cref{fig:fvfangres}, and the gray to those in \cref{fig:bvmangres}. The right panel shows distributions of $\delta \Psi$ in two different energy slices, between \SIrange{10}{100}{\tera \eV} (top) and \SIrange{1}{10}{\peta \eV} (bottom), with line colors and styles matching those of the left panel.}
\label{fig:dvfangres}
\end{figure}
\Cref{fig:dvfangres} shows the improvement in the angular resolution of the benchmark MC. Distributions of $\delta \Psi$ with the two-cascade, single-cascade and original Mie model are shown in solid blue, dashed orange and dotted gray, respectively. The left panel shows the median angular resolution (solid, dashed and dotted lines) as a function of $\edp$. Again, bands correspond to the \SI{25}{\%} and \SI{75}{\%} intervals in the $\delta \Psi$ distribution. The right panels show the full $\delta \Psi$ distribution in two energy regimes, between \SIrange{10}{100}{\tera \eV} (top) and \SIrange{1}{10}{\peta \eV} (bottom), with line colors and styles matching those of the left panel. We see that including a two-cascade reconstruction leads to some further improvement in the directional reconstruction, since it better describes the longitudinal extension of particle showers. It is worth emphasizing that the corrections described in \crefrange{sec:bspl}{sec:tilt} are prerequisite; a two-cascade reconstruction exhibits improvements only when an accurate single cascade model is used.

\subsection{Ice systematic uncertainties}
\label{sec:syst}

One ice systematic that can affect shower directional reconstruction is the bubble column, or hole ice, that formed as part of the drill-hole refreezing process~\cite{IceCube:2023ahv}. Based on camera footage and in situ calibration data, the hole ice is known as a centrally located region of heightened scattering and absorption. As its optical properties are less understood than the bulk ice, it has traditionally been modeled as a global modification in the DOM acceptance as a function of the incident photon direction~\cite{IceCube:2023ahv}. When the forward scattering region is strongly modified, a degradation of the angular resolution is observed on the order of \SIrange{0.5}{1}{\degree}. As the angular sensitivity curves modify photon acceptance along the (downward-facing) PMT axis, mismodeling of the hole ice can also pull the reconstructed $\hatzen$ by up to \SIrange{2}{3}{\degree} for events arriving horizontally, with smaller pulls for non-horizontal events.

The optimal results obtained in this work, shown in \cref{fig:dvfangres}, rely on B-spline surfaces fitted to an ice model similar to the one used in the benchmark simulation; the hole ice model is identical, and only minor differences---in bulk ice optical properties and layer undulations---exist between the model used to construct \cref{eq:full} and that used in the MC. Exclusive of hole ice, a variance of the other ice optical properties at the percent level, which is on the order of the current uncertainty envelope, was found to have a negligible impact on angular resolution.

Further, there are effects for which robust quantification of uncertainty currently does not exist, such as birefringence and the extrapolated layer undulation. Given the accuracy with which calibration data is now described~\cite{tc-18-75-2024}, any residual systematic mismodeling of the anisotropy along the ice flow axis is likely to be much smaller than what is shown in \cref{fig:fvbangres}. Mismodeling of ice layer undulations outside the instrumented footprint could also have an impact on the reconstruction of events that occurred outside the detector. These events tend to be more difficult to reconstruct, with worsening resolution as the interaction vertex moves further away from the instrumented region due to increasingly limited arrival photon statistics, and inline with what might be expected of the accuracy of the extrapolation itself. Thus, the extrapolation systematic should be a subdominant effect. Finally, there are potentially undiscovered systematics, the impact of which cannot be evaluated based on simulation alone and may require data-driven approaches to quantify. Such effects lie beyond the scope of this work, though the general techniques described here should be applicable as future improvements are discovered.

\section{Summary}
\label{sec:summary}

Much progress has been made over the past few years towards an improved understanding of the South Pole ice, including a microscopic description of the observed anisotropy along the ice flow axis attributed to polycrystalline birefrigence~\cite{tc-18-75-2024}, and a detailed mapping of ice isochron undulations across the detector~\cite{IceCube:2023qua}. These refinements to the ice model can be incorporated into the reconstruction of in-ice particle showers, either with neural networks~\cite{Abbasi:2021ryj,IceCube:2021umt,IceCube:2023avo} or via a series of physically motivated corrections, as discussed in \cref{sec:bfr} and \cref{sec:tilt}. Using \cref{eq:full}, the shower longitudinal extension can be approximated with a two-cascade model, as highlighted in \cref{sec:dc}. \Cref{tab:angres} gives an overview of the median angular resolutions obtained for our benchmark MC, which is simulated with recent updates to ice modeling that include birefringence~\cite{tc-18-75-2024} and ice layer undulations~\cite{IceCube:2023qua}, at four different energies as cumulative model improvements are included in reconstruction. At energies above \SI{1}{\peta \eV}, a \SI{3.5}{\degree} median angular resolution is achieved when all corrections are included, which is over a factor of three improvement compared to using only \cref{eq:bulk}.
\begin{table}[htbp]
\centering
\caption{The median angular resolution at the energies listed for the given models. A detailed description of improvements in model construction can be found in \cref{sec:bspl}, the birefringence correction in \cref{sec:bfr}, the ice layer undulations in \cref{sec:tilt}, and the approximation of shower extension in \cref{sec:dc}. Compared to models that do not include any corrections, the angular resolution is improved by more than a factor of three at \SI{1}{\peta \eV}.\label{tab:angres}}
\smallskip
\begin{tabular}{l|rrrr}
\hline
Corrections applied &\SI{50}{\tera \eV}&\SI{100}{\tera \eV}&\SI{1}{\peta \eV}&\SI{5}{\peta \eV}\\
\hline
None (Mie model) & $11.0^\circ$ & $11.9^\circ$ & $12.0^\circ$ & $12.6^\circ$\\
None & $10.7^\circ$ & $11.3^\circ$ & $11.0^\circ$ & $11.0^\circ$\\
Birefringence & $6.1^\circ$ & $5.8^\circ$ & $5.7^\circ$ & $5.7^\circ$\\
Birefringence + undulations & $4.6^\circ$ & $4.4^\circ$ & $4.0^\circ$ & $3.9^\circ$\\
Birefringence + undulations + shower extension & $4.4^\circ$ & $4.2^\circ$ & $3.5^\circ$ & $3.5^\circ$ \\
\hline
\end{tabular}
\end{table}

Ice model systematics, in particular bubble columns that formed as part of the drill-hole refreezing process~\cite{IceCube:2023ahv}, can lead to some degradation in resolution as discussed in \cref{sec:syst}. For the variations tested, a worsening in the median angular resolution of up to \SI{1}{\degree} is seen when differences in the forward-scattering region are large. A bias in the reconstructed zenith distribution is also observed. Note, however, that with smoother B-spline surfaces, the more artificial zenith biases have been mitigated using the methods discussed in \cref{sec:bspl}. Additionally, the IceCube Upgrade, which will be a dense infill extension at the center of IceCube, is currently planned for installation in the next couple years~\cite{Ishihara:2019aao}. In addition to new optical modules, next-generation calibration devices will be deployed that should provide additional information about the optical properties of the bubble columns~\cite{Rongen:2021rgc}.

\begin{figure}[hbt]
\centering
\includegraphics[width=0.49\linewidth]{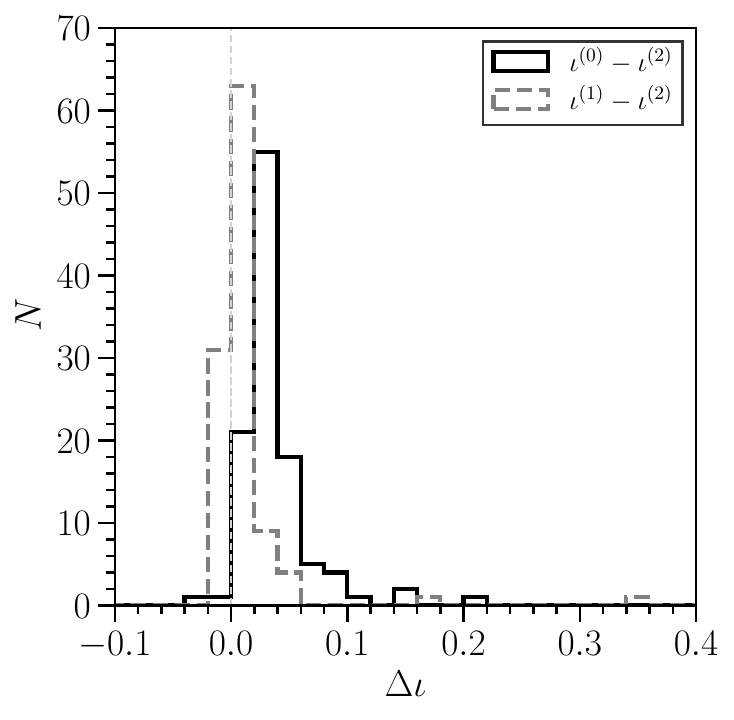}
\includegraphics[width=0.49\linewidth]{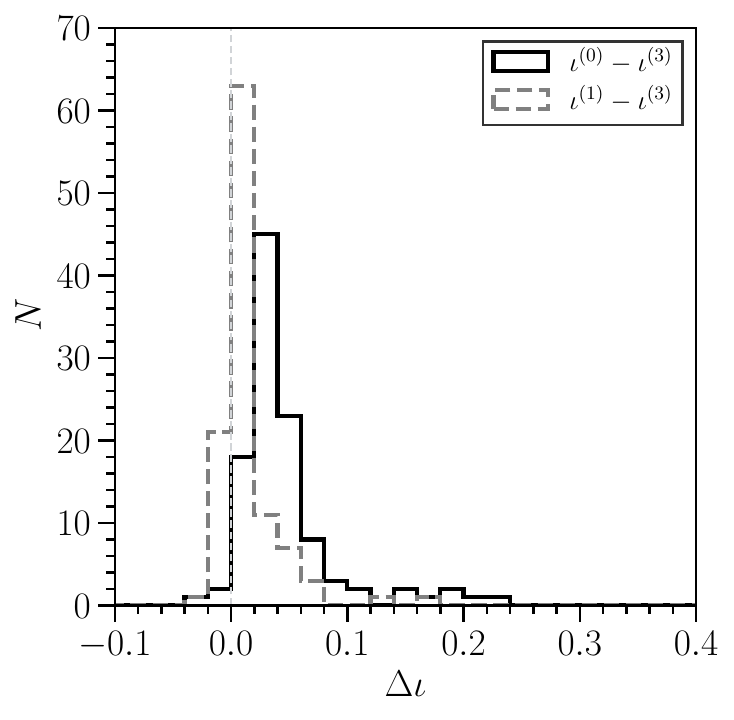}
\caption{Distributions of $\Delta \iota$ for a sample of HESE data events that were reconstructed as showers in Ref.~\cite{IceCube:2023sov}. The left panel shows a comparison similar to the one in \cref{fig:iota}, where the reduced negative log-likelihood for a model that includes corrections due to both layer undulations and birefringence is given by $\iot{2}$. Turning off the layer undulation correction yields $\iot{1}$, and turning off all corrections $\iot{0}$. Since smaller $\iota$ values corresponds to improved agreement against data, the improved description with $\iot{2}$ is evidenced by the skew towards positive values. The right panel shows a similar comparison but changing the reference to $\iot{3}$, which is obtained with the additional two-cascade approximation as described in \cref{sec:dc}. In both panels, the vertical dashed line at $\Delta \iota=0$ is a visual guide to highlight the positive skew.}
\label{fig:diota}
\end{figure}
As a final comparison, a check on data was performed using 109 HESE events that were reconstructed as showers in Ref.~\cite{IceCube:2023sov}. \Cref{fig:diota} shows improvements in the reduced negative log-likelihood, $\iota$, similar to that shown for the benchmark MC in \cref{fig:iota}. The left panel shows $\Delta \iota$ distributions relative to $\iot{2}$, where $\iot{2}$ is obtained using the best-fit from \cref{eq:full}, $\iot{1}$ using \cref{eq:flat}, and $\iot{0}$ using \cref{eq:bulk}, all with the fixes described in \cref{sec:bspl}. As smaller $\iota$ values correspond to a better description of data, the positive skews in the $\Delta \iota$ distributions illustrate the improvement in describing data when using \cref{eq:full}. The skew is larger for $\iot{0} - \iot{2}$ than for $\iot{1} - \iot{2}$ because \cref{eq:flat} includes the birefringence correction. The right panel shows similar comparisons except now relative to $\iot{3}$, which is the best-fit obtained using \cref{eq:full} and the two-cascade approximation. We see that $\Delta \iota$ shifts to more positive values, indicating that the two-cascade model yields a further improvement in the description of data.

While it may seem that increases in ice model complexity pose a challenge for the accurate reconstruction of particle showers in IceCube, in reality the problem can be naturally broken down into a few pieces, thus restoring some symmetry and reducing dimensionality. Birefringence is a global property that can be approximated with a few coordinate transformations, while ice layer undulations can be approximated by shifting the shower's physical depth to its corresponding $\Pref$-depth. The shower profile itself can then be approximated using a two-cascade model. Implementing these features leads to improvement in the angular resolution by more than a factor of three. Hopefully, this work will aid future analyses and inform technical progress towards reaching the statistical limit for particle shower reconstruction in the South Pole ice.

\acknowledgments

The IceCube Collaboration acknowledges the significant contributions to this manuscript from Tianlu Yuan. The authors gratefully acknowledge the support from the following agencies and institutions: 
USA {\textendash} U.S. National Science Foundation-Office of Polar Programs,
U.S. National Science Foundation-Physics Division,
U.S. National Science Foundation-EPSCoR,
U.S. National Science Foundation-Office of Advanced Cyberinfrastructure,
Wisconsin Alumni Research Foundation,
Center for High Throughput Computing (CHTC) at the University of Wisconsin{\textendash}Madison,
Open Science Grid (OSG),
Partnership to Advance Throughput Computing (PATh),
Advanced Cyberinfrastructure Coordination Ecosystem: Services {\&} Support (ACCESS),
Frontera computing project at the Texas Advanced Computing Center,
U.S. Department of Energy-National Energy Research Scientific Computing Center,
Particle astrophysics research computing center at the University of Maryland,
Institute for Cyber-Enabled Research at Michigan State University,
Astroparticle physics computational facility at Marquette University,
NVIDIA Corporation,
and Google Cloud Platform;
Belgium {\textendash} Funds for Scientific Research (FRS-FNRS and FWO),
FWO Odysseus and Big Science programmes,
and Belgian Federal Science Policy Office (Belspo);
Germany {\textendash} Bundesministerium f{\"u}r Bildung und Forschung (BMBF),
Deutsche Forschungsgemeinschaft (DFG),
Helmholtz Alliance for Astroparticle Physics (HAP),
Initiative and Networking Fund of the Helmholtz Association,
Deutsches Elektronen Synchrotron (DESY),
and High Performance Computing cluster of the RWTH Aachen;
Sweden {\textendash} Swedish Research Council,
Swedish Polar Research Secretariat,
Swedish National Infrastructure for Computing (SNIC),
and Knut and Alice Wallenberg Foundation;
European Union {\textendash} EGI Advanced Computing for research;
Australia {\textendash} Australian Research Council;
Canada {\textendash} Natural Sciences and Engineering Research Council of Canada,
Calcul Qu{\'e}bec, Compute Ontario, Canada Foundation for Innovation, WestGrid, and Digital Research Alliance of Canada;
Denmark {\textendash} Villum Fonden, Carlsberg Foundation, and European Commission;
New Zealand {\textendash} Marsden Fund;
Japan {\textendash} Japan Society for Promotion of Science (JSPS)
and Institute for Global Prominent Research (IGPR) of Chiba University;
Korea {\textendash} National Research Foundation of Korea (NRF);
Switzerland {\textendash} Swiss National Science Foundation (SNSF).


\bibliographystyle{JHEP}
\bibliography{biblio.bib}


\end{document}

%% file: authors.tex
\author[16]{R. Abbasi,}
\author[63]{M. Ackermann,}
\author[17]{J. Adams,}
\author[39,a]{S. K. Agarwalla,}
\author[11]{J. A. Aguilar,}
\author[21]{M. Ahlers,}
\author[22]{J.M. Alameddine,}
\author[43]{N. M. Amin,}
\author[41]{K. Andeen,}
\author[25]{G. Anton,}
% ...
\affiliation[11]{Universit{\'e} Libre de Bruxelles, Science Faculty CP230, B-1050 Brussels, Belgium}
% ...
\affiliation[16]{Department of Physics, Loyola University Chicago, Chicago, IL 60660, USA}
\affiliation[17]{Dept. of Physics and Astronomy, University of Canterbury, Private Bag 4800, Christchurch, New Zealand}
% ...
\affiliation[21]{Niels Bohr Institute, University of Copenhagen, DK-2100 Copenhagen, Denmark}
\affiliation[22]{Dept. of Physics, TU Dortmund University, D-44221 Dortmund, Germany}
% ...
\affiliation[25]{Erlangen Centre for Astroparticle Physics, Friedrich-Alexander-Universit{\"a}t Erlangen-N{\"u}rnberg, D-91058 Erlangen, Germany}
% ...
\affiliation[39]{Dept. of Physics and Wisconsin IceCube Particle Astrophysics Center, University of Wisconsin{\textemdash}Madison, Madison, WI 53706, USA}
% ...
\affiliation[41]{Department of Physics, Marquette University, Milwaukee, WI 53201, USA}
% ...
\affiliation[43]{Bartol Research Institute and Dept. of Physics and Astronomy, University of Delaware, Newark, DE 19716, USA}
% ...
\affiliation[63]{Deutsches Elektronen-Synchrotron DESY, Platanenallee 6, D-15738 Zeuthen, Germany}
\affiliation[a]{also at Institute of Physics, Sachivalaya Marg, Sainik School Post, Bhubaneswar 751005, India}
% ...
